%% file: main.tex
\documentclass{article}

\usepackage{arxiv}
\usepackage[utf8]{inputenc}
\usepackage[T1]{fontenc}
\usepackage{microtype}
\usepackage{inconsolata}
\usepackage{graphicx}
\usepackage{amsthm}
\usepackage{amsmath,amssymb,amsfonts}
\usepackage{subcaption}
\usepackage{textcomp}
\usepackage{xcolor}
\usepackage{xspace}
\usepackage{tcolorbox}
\usepackage{multirow}
\usepackage{colortbl}
\usepackage{fixmath}
\usepackage{hyperref}
\usepackage{enumitem}
\usepackage{algorithmicx,algorithm,algpseudocode}
\usepackage{booktabs}
\usepackage{url}
\usepackage{soul}
\usepackage{tabularx, diagbox}
\usepackage{comment}
\usepackage{wrapfig}
\usepackage{natbib}

\definecolor{L0}{RGB}{0,0,0}
\definecolor{L1}{RGB}{210,2,0}
\definecolor{L2}{RGB}{245,197,3}
\definecolor{L3}{RGB}{249,153,53}
\definecolor{L4}{RGB}{0,160,234}
\definecolor{L5}{RGB}{37,172,116}
\definecolor{L6}{RGB}{139,9,135}
\definecolor{L7}{RGB}{215,160,80}
\definecolor{L8}{RGB}{229,223,135}
\definecolor{L9}{RGB}{130,140,71}
\definecolor{L10}{RGB}{186,101,81}
\definecolor{L12}{RGB}{25,99,164}
\definecolor{L14}{RGB}{178,137,195}
\definecolor{L17}{RGB}{12,68,85}

\newcommand{\EM}[1]{\textbf{\emph{#1}}}

\newcommand{\SOTA}[1]{\textbf{#1}}
\newcommand{\hide}[1]{}

\theoremstyle{plain}
\newtheorem{theorem}{Theorem}

\theoremstyle{definition}
\newtheorem{definition}{Definition}

\theoremstyle{remark}

\newcolumntype{C}{>{\centering\arraybackslash}X}
\newcommand{\name}{{\textsc{CodeRefuser}}\xspace}

\title{Task Abstention for Large Language Models \\ in Code Generation}

\author{}

\makeatletter
\newcommand{\njuaffil}{%
  State Key Laboratory of\\
  Novel Software Technology\\
  Nanjing University}
\newcommand{\lonaffil}{%
  School of Computing and\\
  Mathematical Sciences\\
  University of London}
\newcommand{\authblock}[2]{%
  \begin{minipage}[t]{0.31\textwidth}
    \centering
    \textbf{#1}\\[3pt]
    {\footnotesize #2}%
  \end{minipage}}
\renewcommand{\@maketitle}{%
  \vbox{%
    \hsize\textwidth
    \linewidth\hsize
    \vskip 0.1in
    \@toptitlebar
    \centering
    {\LARGE\sc \@title\par}
    \@bottomtitlebar
    \vskip 0.12in
    \normalsize
    \authblock{Yanke Zhou}{\njuaffil}\hfill
    \authblock{Yuhao Tan}{\njuaffil}\hfill
    \authblock{Senrong Xu}{\njuaffil}\par
    \vskip 7pt
    \authblock{Zenan Li}{\njuaffil}\hfill
    \authblock{Yuan Yao}{\njuaffil}\hfill
    \authblock{Taolue Chen}{\lonaffil}\par
    \vskip 7pt
    \begin{minipage}[t]{0.31\textwidth}\strut\end{minipage}\hfill
    \authblock{Xiaoxing Ma}{\njuaffil}\hfill
    \begin{minipage}[t]{0.31\textwidth}\strut\end{minipage}\par
    \vskip 0.15in
  }%
}
\makeatother

\begin{document}
\maketitle
\begingroup
\renewcommand{\thefootnote}{}%
\footnotetext{\textbf{E-mail}:~yankezhou@smail.nju.edu.cn}%
\endgroup

\begin{abstract}
Large language models (LLMs) have revolutionized automated code generation.
One serious concern, however, is the so-called ``hallucination'', i.e., LLMs may generate seemingly plausible but functionally incorrect code. In this paper, we study the task abstention problem, i.e., determining whether a given LLM should abstain from performing a specific code generation task to avoid likely hallucination. Our approach features a calibrated abstention rule, grounded in the principles of multiple hypothesis testing. The rule assesses generation consistency through code execution outcomes, allowing it to handle syntactic diversity of semantically equivalent code without reliance on oracle test cases or external databases.
We prove that our approach provides a rigorous, distribution-free theoretical guarantee on its abstention decisions.
We evaluate our method on benchmark datasets using several open-source code LLMs. Results show that our method allows generative models to more accurately and efficiently identify and abstain from tasks that induce hallucination compared to existing techniques, providing a reliable mechanism for safer and more robust code generation.
\end{abstract}

\input{src/introduction}
\input{src/problem}

\input{src/method}

\input{src/experiment}
\input{src/conclusion}

\section*{Limitations}

\noindent{\em Generalization to out-of-distribution tasks.}  Our theoretical guarantee relies on the assumption that the calibration and test data are independent and identically distributed. However, in real-world scenarios, the problem a system encounters may differ from those it sees during calibration. This potential distribution shift could introduce bias and weaken the risk control guarantees in practice.
To mitigate this threat, we have conducted experiments using MBPP as calibration data and HumanEval as test data, and the results show that there was no significant decline in performance.

\smallskip
\noindent{\em Applicability to non-deterministic systems.} The core of our approach--both the score functions and the STDF mechanism---assumes that all tasks have deterministic solutions.
The assumption may not hold for complex, interactive software systems, which may exhibit nondeterminism. As our method relies on clustering identical outputs, its applicability is currently limited to deterministic problem domains.

\smallskip
\noindent{\em Evaluation limited to a single programming language.} Current experiments are conducted in Python, a language where modern LLMs have demonstrated strong performance~\cite{huang2024effibench, twist2025llms}. However, many real-world systems are built using other languages such as Java or C, which may have different characteristics
(e.g., pointer and memory management in C). The effectiveness of our method
in these programming languages
requires further evaluation.

\bibliographystyle{unsrtnat}
\bibliography{custom}

\appendix
\input{src/related_work}
\input{src/LTT}
\input{tables/transfer_pic}
\input{src/scoredef}
\input{src/STDF}
\input{src/extra_eval}

\end{document}

%% file: src/introduction.tex
\section{Introduction} \label{sect:intro}
The recent advancements in large language models (LLMs) are revolutionizing the field of code generation~\cite{alphacodeCompetition-level, chenEvaluatingLargeLanguage2021, roziere2023code}.
As these models are increasingly integrated into software development workflows, ensuring their trustworthiness and reliability has become a pressing requirement.
However, current LLMs tend to hallucinate, i.e., they may produce outputs that are seemingly plausible but are actually incorrect~\cite{Huang_2025}. This paper is particularly concerned with mitigating hallucination in the context of code generation, for which ~\citet{lee2025hallucination} provided a recent survey.

Existing work on code hallucination detection has predominantly focused on \emph{sample hallucination}, that is, a generated code snippet that fails to execute as expected or meet specified requirements, despite being syntactically correct or even semantically plausible~\cite{tian2025codehalu}. 
However, there is an arguably more fundamental source of hallucination, i.e., the task itself.
This might be caused by ambiguous or unclear prompts, or by the inherent limitation of current LLMs which are doomed to fail on certain problems.
To this end, we propose to study the {\em task abstention} problem for code generation, which aims to faithfully detect code generation tasks that an LLM is unlikely to solve.

\hide{
For example, consider the code generation task of ``appending a new entry to a \texttt{pandas.DataFrame}.'' 
The \texttt{DataFrame.append} method was deprecated after pandas version 2.0 was released and thus superseded by the \texttt{pandas.concat} function. However, an LLM trained on the outdated corpora may consistently generate solutions using the deprecated method. Such code will invariably fail when executed in any up-to-date environment. 
In such a scenario, sample-level hallucination mitigation techniques are rendered ineffective, because the root issue is not an isolated flaw in a single generated sample but a fundamental, outdated understanding of correctly performing the task.}

\hide{
In this paper, we propose to study hallucination at the task level, which refers to cases where a model is consistently unable to generate correct solutions for a specific code generation task. In particular, we formulate the {\em task abstention} problem, which aims to faithfully detect code generation tasks that an LLM is unlikely to be able to solve.}

To be specific, we formulate the basic concept of task abstention and propose a comprehensive approach, \name, to control its associated risks with theoretical guarantees. 
Our approach is built upon the LTT framework~\cite{angelopoulos2021learn} and consists of two phases: {\em calibration} and {\em testing}.  
During the calibration phase, different from traditional NLP tasks, we propose to use test cases when constructing the calibration set. The intuition is that syntactically different programs may exhibit the same semantics, and the correctness of programs should be better determined by running the test cases. 
Scoring function lies at the heart of the LTT framework. We then define two score functions leveraging the execution-based results. One issue here is that oracle test cases are not usually available during the testing phase. To this end, we prompt the LLM to generate not only code solutions but also corresponding test cases, and propose a sample-test dual filtering mechanism to deal with potentially flawed, model-generated test cases.

We conduct evaluations across several representative code LLMs and code generation benchmarks. 
The results demonstrate the effectiveness of our method on the task abstention problem for LLM-based code generation. For example, on average, our method achieves 26.5\% absolute improvement compared to the best existing competitor in terms of abstention precision. By analyzing the results, we find that: 1) static methods that adopt traditional NLP score functions are insufficient for accurate abstention; and 2) running the generated code is necessary but heavily relies on the quality of generated tests, and our sample-test dual filtering mechanism is crucial for the effectiveness.
Additionally, the theoretical guarantees are confirmed by our empirical experiments. 

Our contributions can be summarized as follows. (1) We introduce and formalize the concept of {task abstention} in the context of LLM-based code generation, in contrast to the {sample-level} hallucination. (2) We propose a novel approach for accurate {task abstention}, enabling the model to answer ``I don't know'' when encountering tasks it is unlikely to solve. The approach features two advantages: i) {\em rigorous theoretical guarantees} grounded in multiple hypothesis testing, and ii) {\em a minimal reliance on the oracle test cases or external references}. 


%% file: src/problem.tex
\section{Problem Formulation} \label{sect:prel}
Throughout the paper, we use $\mathcal{X}$ to stand for the input space consisting of a set of tasks (i.e., prompts) for LLMs, and $\mathcal{Y}$ for the output space. Typically, $y\in \mathcal{Y}$ is a code snippet generated by the LLM. The task abstention problem is defined as follows.

\begin{definition}[Task Abstention for LLM-based Code Generation]\label{def:problem}
    Consider a code generation task (prompt) $x \in \mathcal{X}$ and an LLM $\mathcal{M}: \mathcal{X} \rightarrow \mathcal{Y}$ which generates a code snippet $y \in \mathcal{Y}$. 
    Task abstention aims to find a \emph{refusal function} $r: \mathcal{X} \rightarrow \{0,1\}$ for $\mathcal{M}$, where $r(x) = 1$ indicates $\mathcal{M}$ should abstain from answering and $r(x)=0$ indicates otherwise.
\end{definition}

\noindent {\em When to refuse the task.} 
In this work, for code hallucination we adopt the definition of Tian et al.~\cite{tian2025codehalu}, considering a plausible but functionally incorrect code snippet as hallucination, although the proposed approach can be adapted to handle other types of hallucination~\cite{lee2025hallucination}. As a result, a desirable refusal function $r$ should be able to identify the case when the given LLM is unlikely to produce a functionally correct code snippet for the given task.

\smallskip 
\noindent  {\em Task abstention vs. sample hallucination detection.} 
Sample hallucination detection aims to detect {\em individual} code snippets generated by the model that cannot be executed as expected.
In contrast, task abstention aims to identify the prompts for which an LLM is unlikely to generate the correct answer, 
even if the LLM is allowed to produce a high volume of samples.

\smallskip
\noindent{\em Evaluation Criterion.} 
We instantiate the refusal function by defining a criterion. 
Specifically, for a given prompt $x$, $\mathcal{M}(x)$ gives a distribution on the output space $\mathcal{Y}$, and thus a random output is (functionally) correct for a certain probability (which is unknown in general). We define a new metric $H@k$, which is the probability that $k$ randomly generated samples are all \emph{incorrect}.

Ideally, when the oracle test cases are available, we can determine if a generated code sample is correct by executing these test cases. We consider the process of first generating a sample of $n$ ($n>k$) instances, and then count the number of incorrect samples. Assuming that these $n$ generated samples contain $c$ correct instances, $H@k$ metric can be estimated by 
\begin{equation} \label{eq:h_k}
    H@k := \frac{\binom{n-c}{k}}{\binom{n}{k}} .
\end{equation}

\begin{definition}[$(k, \alpha)$-Criterion for Task Abstention] \label{def:abstain}
    The given LLM should refuse the code generation task if its $H@k$ metric exceeds a threshold $\alpha$, i.e., $H@k > \alpha$.
\end{definition}

Henceforth $\alpha$ is referred to as the {\em risk tolerance}, which is fixed in advance. For example, $H@k > 0.8$ means that the LLM cannot generate a correct code snippet in $k$ attempts with probability greater than 0.8.

%% file: src/method.tex
\begin{figure*}[htbp]
    \centering
    \includegraphics[width=\linewidth]{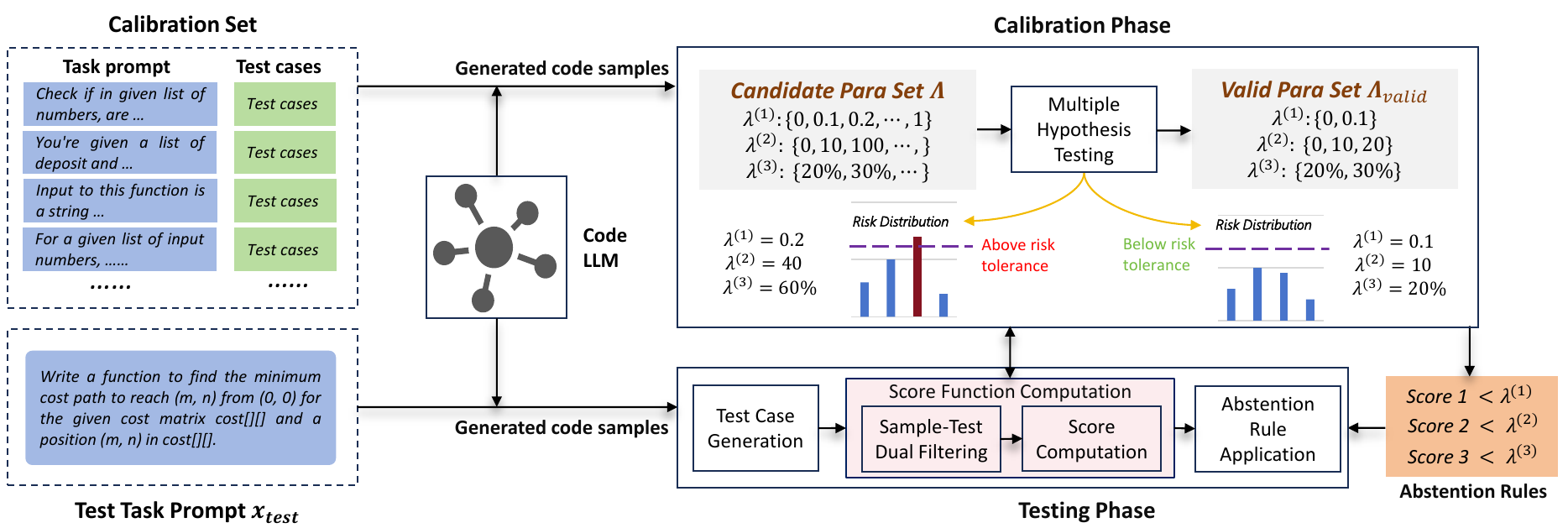}
    \caption{The Overview of \name.}
    \label{fig:arch}
\end{figure*}

\section{Methodology} \label{sect:method}

\subsection{Overview} 
In this work, we build our task abstention approach upon the Learn Then Test (LTT) framework~\cite{angelopoulos2021learn}. We choose LTT as it can provide
statistical guarantees for machine learning models by simply adding
a post-processing step after the model is trained.
Specifically, LTT allows us to calibrate a threshold $\lambda$ using a calibration set $\mathcal{D}_{cal}$; 
it guarantees that for a new test task $x_{test}$, the risk of accepting an incompetent task is controlled with high probability:
\begin{equation} \label{eq:LTT}
\mathbb{P}\Big(\mathbb{E}[R(x_{test}; \hat{\lambda}) \mid \mathcal{D}_{cal}] \leq \alpha\Big) \geq 1 - \delta,
\end{equation}
where $R$ is the admission risk (discussed later in Section~\ref{sec:cali}), $\alpha$ is the user-specified risk tolerance, and $1-\delta$ is the confidence level (e.g., 90\%). We refer readers to Appendix~\ref{appd:ltt} for the theoretical preliminaries of LTT and the multiple hypothesis testing procedure used to derive $\hat{\lambda}$.

An overview of the proposed approach \name is depicted in Fig.~\ref{fig:arch}, which is divided into {\em calibration phase} and {\em testing phase}. 

\EM{The calibration phase} takes as input a calibration set, which contains a set of task descriptions and the corresponding test cases. The output of the calibration phase is a set of abstention rules. Specifically, following the LTT framework, \name first defines an {\em admission risk} measuring the risk that the LLM cannot generate the correct code for an admitted task. Then, \name defines a set of {\em score functions}, and derives the computed scores for each task in the calibration set. For each score function, a corresponding abstention rule is obtained, by selecting the suitable parameters (i.e., a set of valid thresholds, $\Lambda_{valid}$) so that the admission risk is below a given tolerance level (i.e., $\alpha$), under multiple hypothesis testing with a given significance level (i.e., $\delta$). The {\em abstention rules} are in the form of a conjunction of inequalities, each describing the relationship between a score function applied on the input code generation task and the corresponding valid threshold.

\EM{The testing phase} takes the current code generation task as input, and determines whether the LLM should abstain from generating the code. For any given test task (prompt), \name first generates a set of test cases, based on which the score functions can be applied. A key issue is that the generated test cases may be invalid. To mitigate this issue, we propose a {\em sample-test dual filtering} mechanism. \name then applies the abstention rules on the test task, based on which the abstention decision is made.

Note that
the calibration phase can be done in advance. Once the abstention rules are obtained, we can guarantee that, during the testing phase when an i.i.d. code generation task is given,\footnote{One may argue that the i.i.d. assumption, which is essential for the statistical guarantee, is too strong in practice. However, as will be shown in the experiments later, we observe that the abstention rules obtained from one dataset transfer well when they are tested on other datasets.} the performance of \name is theoretically guaranteed by Eq.~\eqref{eq:LTT}, i.e., the LLM ensures that once the code generation task is admitted, the risk of failing the task is below the tolerance $\alpha$ with probability at least $1-\delta$.

\subsection{The Calibration Phase}\label{sec:cali}
The calibration phase consists of several key components, including the calibration set construction, the admission risk definition and evaluation, the determination of valid threshold set, and the definition of score functions. We mainly describe the former three components here, and leave the score functions in Section~\ref{sec:3.3}, as they will be reused by the testing phase.

\smallskip 
\noindent{\bf Calibration Set Construction}.
In code generation, a key difference from the standard LTT framework lies in the output space, which is generally infinite. 
As a result, instead of providing the code generation prompt and the corresponding correct code as the calibration set, we use the code generation prompt and the corresponding oracle test cases.
That is, 
$\mathcal{D}_{\rm cal} = \lbrace (x_i, t_i) \rbrace_{i=1}^m$.
In general, the correctness of the generated code snippet $y_i$ for $x_i$ is determined by executing $y_i$ on the test cases $t_i$.

\smallskip 
\noindent{\bf Admission Risk}.
Recall that our goal is to ensure that model $\mathcal{M}$'s risk on any given task $x$ is below the risk tolerance $\alpha$.
We refer to this risk as admission risk, and define it as 
\begin{equation} \label{eq:loss}
    R(x) = (1 - r(x)) \cdot H@k,
\end{equation}
where $r(x)$ is the refusal function in Definition~\ref{def:problem}, and $H@k$ is defined in Eq.~\eqref{eq:h_k}. 
Essentially, Eq.~\eqref{eq:loss} measures the risk when a code generation task is admitted (i.e., $r(x)=0$), but the LLM cannot generate the correct code with probability $H@k$ in $k$ attempts.

During the calibration phase, since we have the oracle test cases, $H@k$ can be accurately evaluated. For $r(x)$, we also resort to the calibration set and design abstention rules parameterized by the parameters $\pmb{\lambda} = (\lambda_1,\cdots, \lambda_N)$ as follows: 
\begin{equation}
    \hat{r}(x; \pmb{\lambda}) = 
    \begin{cases}
        1,  &   \mbox{if } \forall 1\leq i\leq N.\ g_i(x) < \lambda_i, \\
        0,  &   \mbox{o.w.}
    \end{cases}
\end{equation}
where each $g_i(x) \in \mathbb{R}$ yields a score for the current code generation task $x$, and $\pmb{g} = (g_1,\cdots g_N)$ represents the vector of such score functions. In the calibration phase, the computed scores are used to determine the valid thresholds, so that the admission risk is controlled below the pre-specified risk tolerance $\alpha$ with a high probability. Specifically, given the score functions, the choice of $\pmb{\lambda}$ dictates $\hat{r}$, which in turn influences the admission risk, denoted as $\hat{R}(x; \pmb{\lambda})$. 

\smallskip 
\noindent{\bf Determining $\Lambda_{valid}$}. 
Given candidate parameters $\pmb{\lambda}$, the admission risk for each sample $(x_i, t_i) \in \mathcal{D}_{cal}$ is computed as follows. For the prompt $x_i$, we first use the current LLM to generate $n$ code samples. Next, we identify the number of correct samples, $c$, by applying the test cases. Then $H@k$ can be calculated by Eq.~(\ref{eq:h_k}). Once the risk values have been computed across the calibration set for the entire range of candidate parameters, we apply the standard multiple hypothesis testing procedure to identify the set $\Lambda_{valid}$ of valid thresholds.

Let $x_{test}$ be a new test prompt; for the given $\delta \in (0,1)$, if $\pmb{\lambda} \in \Lambda_{valid}$, our abstention rule can determine whether to abstain on $x_{test}$, satisfying the following guarantee,
\begin{equation} \label{eq: pac}
    \mathbb{P}\Big( \hat{R}(x_{test}; \pmb{\lambda}) \le \alpha \Big) \ge 1 - \delta.
\end{equation}
Eq.~\eqref{eq: pac} states that for any task from the test set, our abstention rule provides the following probabilistic guarantee: with a confidence of at least $1 - \delta$, the model will either admit the task, in which case $H@k$ is controlled to be below $\alpha$, or it will abstain. This guarantee directly satisfies the criterion for task abstention established in Definition~ \ref{def:abstain}.


\subsection{Score Function} \label{sec:3.3}
The score function plays a vital role in the LTT performance~\cite{angelopoulos2021learn}. While code generation is a type of generative task, code is a unique modality where code samples with identical semantic meaning can possess vastly different syntactic forms. Consequently, we argue that existing score functions operating at a linguistic level~\cite{quach2023conformal} or a semantic reasoning level~\cite{manakul2023selfcheckgpt} are unsuitable for the code generation task. We therefore propose a scoring scheme based on {\em runtime detection}, which leverages generated test cases.

Specifically, for a given task $x$, the generated code samples $Y=\lbrace y_1,  \cdots, y_n\rbrace$ and a set of test cases $T=\lbrace t_1, \cdots, t_l \rbrace$, we cluster the code samples in $Y$ based on test cases, resulting in a partition $Y= C_1\uplus\cdots \uplus C_h$ such that all code samples from the same cluster yield identical outputs for every test case $t \in T$.\footnote{Note that we use both the test cases in the calibration set and the LLM-generated test cases during calibration, and only LLM-generated test cases in the testing phase. More details can be found in Section~\ref{sect:dual}.}
%
%
Clustering based on execution outputs provides an intrinsic measure of semantic equivalence among code samples. Leveraging these execution-based clusters, we employ two complementary score functions: \EM{confidence-based score}, which measures the degree to which a code sample's semantic behavior is supported by other generated samples (i.e., the size of its cluster relative to $n$)~\cite{chen2022codetcodegenerationgenerated}; \EM{semantic entropy-based score}, which quantifies the overall uncertainty of the task by measuring the diversity of the resulting cluster distribution~\cite{kuhn2023semantic}. Both metrics are rooted in semantic equivalence but serve different granularities (sample-level vs. task-level). The formal definitions and mathematical formulations for these score functions are detailed in Appendix~\ref{appd:score_functions}.

\subsubsection{Sample-Test Dual Filtering}\label{sect:dual}
The above score functions rely on clustering based on LLM-generated test cases. A key issue in this approach is that the generated test cases may contain invalid inputs. Executing code samples on these invalid inputs may lead to undefined behavior. As a result, two semantically identical code samples could be assigned to different clusters.

As an example, consider the task of generating the $n$-th Fibonacci number. By default, a valid input requires $n \ge 0$ for this task, making any negative input invalid. The problem statement does not specify how these invalid inputs should be handled. Now, assume the model generates three solutions that are all functionally correct for valid inputs but handle invalid inputs differently. E.g., one sample might check for negative inputs and raise an \texttt{AssertError}; another could handle all exceptions by returning a default value 0; a third, lacking any specific error handling, might enter an unterminated recursion that triggers   \texttt{RuntimeError}. 

The example illustrates that even when different code samples share consistent logic for a task, their interaction with a diverse (and potentially imperfect) set of generated test cases can expose variations in their behavior. 
This inflates the measure of inconsistency. 
As a result, the abstention rule may be triggered, making LLM refuse to provide an answer, even though the model had already demonstrated its ability to solve the problem correctly.

To address the quality issues inherent in LLM-generated test cases, we propose the \textit{Sample-Test Dual Filtering} (STDF) mechanism. The core of STDF lies in a reciprocal validation process. We rely on semantic consistency to evaluate code quality, but we also leverage the collective behavior of these code samples to audit the test cases.

This design is grounded in the principle that for a deterministic program, valid inputs yield deterministic outputs, whereas invalid inputs trigger undefined behaviors (UB), resulting in highly divergent execution outcomes. 
By detecting test cases that induce high output entropy across the sample population, we can identify and prune invalid inputs. This refined test suite, in turn, eliminates evaluation noise, allowing for a more accurate assessment of the code samples' semantic consistency. 
The mechanism consists of two filtering steps, with algorithm implementation detailed in Appendix~\ref{appd:STDFdetail}.
\begin{enumerate}
    \item {\em Filtering by Error Rate:} Pruning tests that cause widespread execution failures.
    \item {\em Filtering by Output Diversity:} Pruning tests that yield high semantic entropy (indicative of invalid inputs triggering UB).
\end{enumerate}

\subsection{Testing Phase} \label{sect:testing}
When the abstention rules are obtained on the calibration set, in the testing phase, these rules are applied to make the abstention decisions. As outlined in Alg.\ref{alg:UnifiedFilter}, the process begins by generating $n$ code samples $Y$ and test cases $T$ for the input prompt $x$. Crucially, \name applies the STDF mechanism (Line 3) to purge invalid test cases using thresholds $[\lambda_1, \lambda_2, \lambda_3]$. The code samples are then partitioned into semantic clusters $\mathcal{C}$ based on their execution outputs on the remaining valid test cases. The final abstention decision depends on the selected scoring mode.

\smallskip
\noindent \textbf{Cluster Ratio (CR).} This mode enforces model consensus. \name filters the semantic clusters, retaining only those that account for a sufficient proportion of the total samples (i.e., $|C_i|/|Y| \ge \lambda_{score}$). If no cluster meets this confidence threshold (i.e., the filtered set $Y$ becomes empty), the model implies a lack of consensus and \textit{abstains}.

\smallskip
\noindent \textbf{Semantic Entropy (SE).} This mode limits uncertainty. \name calculates the semantic entropy of the cluster distribution. If the entropy exceeds $\lambda_{score}$, indicating high semantic diversity and confusion, the model \textit{abstains}. If the task satisfies the criterion of the chosen mode, it is \textit{admitted}.

\begin{algorithm}[htbp]
\caption{\small \name Inference Procedure}\label{alg:UnifiedFilter}
\begin{algorithmic}[1]
\Require LLM $\mathcal{M}$; prompt $x$; samples $n$; calibrated thresholds $\pmb{\lambda} = [\lambda_1, \lambda_2, \lambda_3, \lambda_{score}]$
\Ensure Abstention decision
\Procedure{CodeRefuser}{$\mathcal{M}, x, n, \pmb{\lambda}$}
    \State $Y, T \gets \text{Generate}(\mathcal{M}, n) $
    \State $T \gets \text{STDF}(Y, T, [\lambda_1, \lambda_2, \lambda_3])$ 
    \State $\mathcal{C} \gets \text{Clustering } Y \text{ by Exec}(Y, T)$
    \If{Mode is CR} 
        \State $Y \gets \bigcup \{C_i \in \mathcal{C} : |C_i|/|Y| \ge \lambda_{score} \}$
            \If{$Y = \emptyset$} \Return \textit{Abstain} \EndIf
        \EndIf
    
    \If{Mode is SE}
        \State \textbf{if} $\text{SE}(\mathcal{C}) > \lambda_{score}$ \textbf{then} \Return \textit{Abstain}
    \EndIf
    \State \Return \textit{Admit}
\EndProcedure
\end{algorithmic}
\end{algorithm}

%% file: src/experiment.tex
\section{Experiments} \label{sect:exp}
In this section, we present the empirical evaluation of \name, focusing on two primary dimensions. First, we assess the \EM{effectiveness} of our approach on the task abstention problem by benchmarking it against existing counterparts. Second, we examine the \EM{theoretical guarantee} to verify whether the risk control promised by the calibration phase is empirically supported.


\subsection{Experimental Setup}
\noindent{\bf Datasets}.
Our experiments are conducted on two standard Python code generation datasets: {\em HumanEval}~\cite{chenEvaluatingLargeLanguage2021} and {\em MBPP}~\cite{austin2021program}. HumanEval consists of 164 hand-crafted programming problems. Each problem is associated with $5-10$ oracle test cases. 
MBPP is a larger dataset of nearly 1,000 problems. Each task is defined by a short description and includes three test cases to verify its correctness. 

\smallskip
\noindent{\bf Evaluation Metrics}.
We use \emph{Precision} and \emph{F1-score} to evaluate the performance on the task abstention problem. To obtain the ground-truth label, we generate 256 samples for each task, and checks them the on the oracle test cases. The abstention label is ture if the $H@k$ is above the given risk tolerance. We deliberately avoid using {\em Recall} as a primary metric due to the following fact. A high recall value can be easily obtained for an abstention rule, by simply refusing most tasks. This is misleading, as it has little practical utility. 

\smallskip
\noindent{\bf Baselines}.
Since no prior work directly addresses task abstention for code generation, we adapt the following baselines for comparison.
\begin{itemize}[leftmargin=*]
    \item \textit{Execution.} This method directly calculates the $H@k$ score using generated test cases and abstains if it exceeds the risk tolerance threshold. It can be seen as an execution-based baseline without adopting the proposed framework.
    \item \textit{PPL}~\cite{huang2023look} and \textit{NLI}~\cite{manakul2023selfcheckgpt}. We integrate these metrics into our LTT framework as alternative score functions. \textit{PPL} uses model perplexity (static), while \textit{NLI} queries the LLM for self-consistency checking.
    \item \textit{CLM}~\cite{quach2023conformal}. This method is originally proposed for natural language generation. We adapt it to our code generation context, and refuse the task if the calibrated valid response set is empty.\footnote{We use the max version of CLM as it demonstrates the best empirical performance.}
    \item \textit{CodeHalu}~\cite{tian2025codehalu}. 
    This is an execution-based sample-level hallucination detector relying on oracle tests. We adapt it by employing LLM-generated tests as pseudo-oracles, and applying the algorithm to identify valid samples and aggregate these predictions to estimate the task-level $H@k$ for the abstention decision.
\end{itemize}
\input{tables/all_result}
\input{tables/static_compare}

\smallskip
\noindent{\bf Implementation Details}.
We conduct the experiments with Python 3.9.19, Pytorch 2.4.0 and vLLM 0.6.1, running on NVIDIA H800 GPUs with CUDA 12.7. We evaluate the performance of the following four code LLMs: Deepseek-Coder-33B~\cite{guo2024deepseek}, Qwen2.5-Coder-32B~\cite{hui2024qwen2}, CodeLlama-7B~\cite{roziere2023code} and WizardCoder-33B-V1.1~\cite{luo2024wizardcoder}. For all these code LLMs, we use a sampling temperature of 0.8 and a top-$p$ of 0.95. 
For each dataset, we randomly allocate 60\% of the data for calibration, and use the rest for testing. We set $\alpha = 0.2, \delta = 0.1, k=3$ by default.

\begin{figure*}[t]
    \centering
    \begin{subfigure}[b]{0.24\textwidth}
        \centering
        \includegraphics[width=\textwidth]{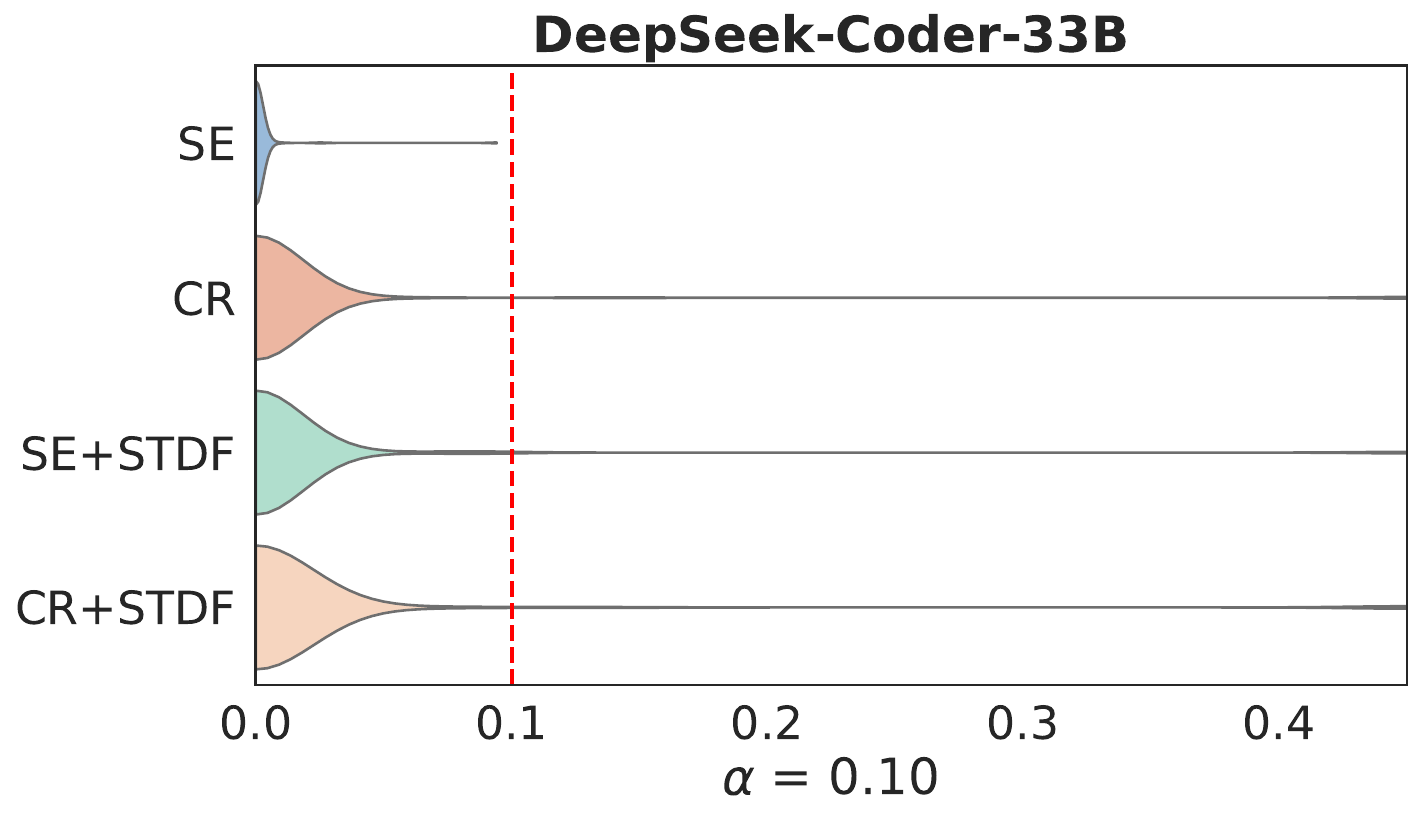}
        \label{fig:first}
    \end{subfigure}
    \hfill
    \begin{subfigure}[b]{0.24\textwidth}
        \centering
        \includegraphics[width=\textwidth]{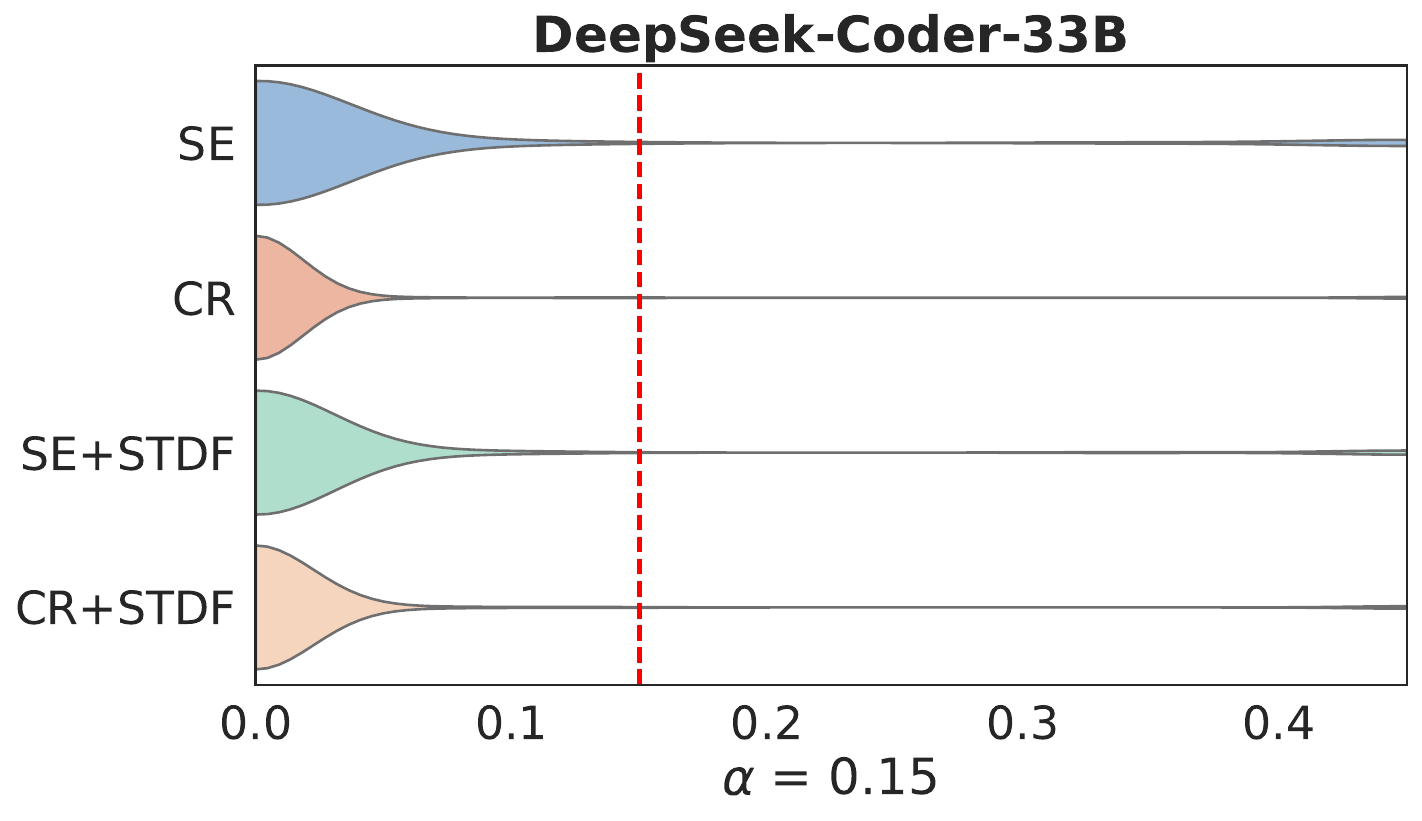}
        \label{fig:second}
    \end{subfigure}
    \hfill
    \begin{subfigure}[b]{0.24\textwidth}
        \centering
        \includegraphics[width=\textwidth]{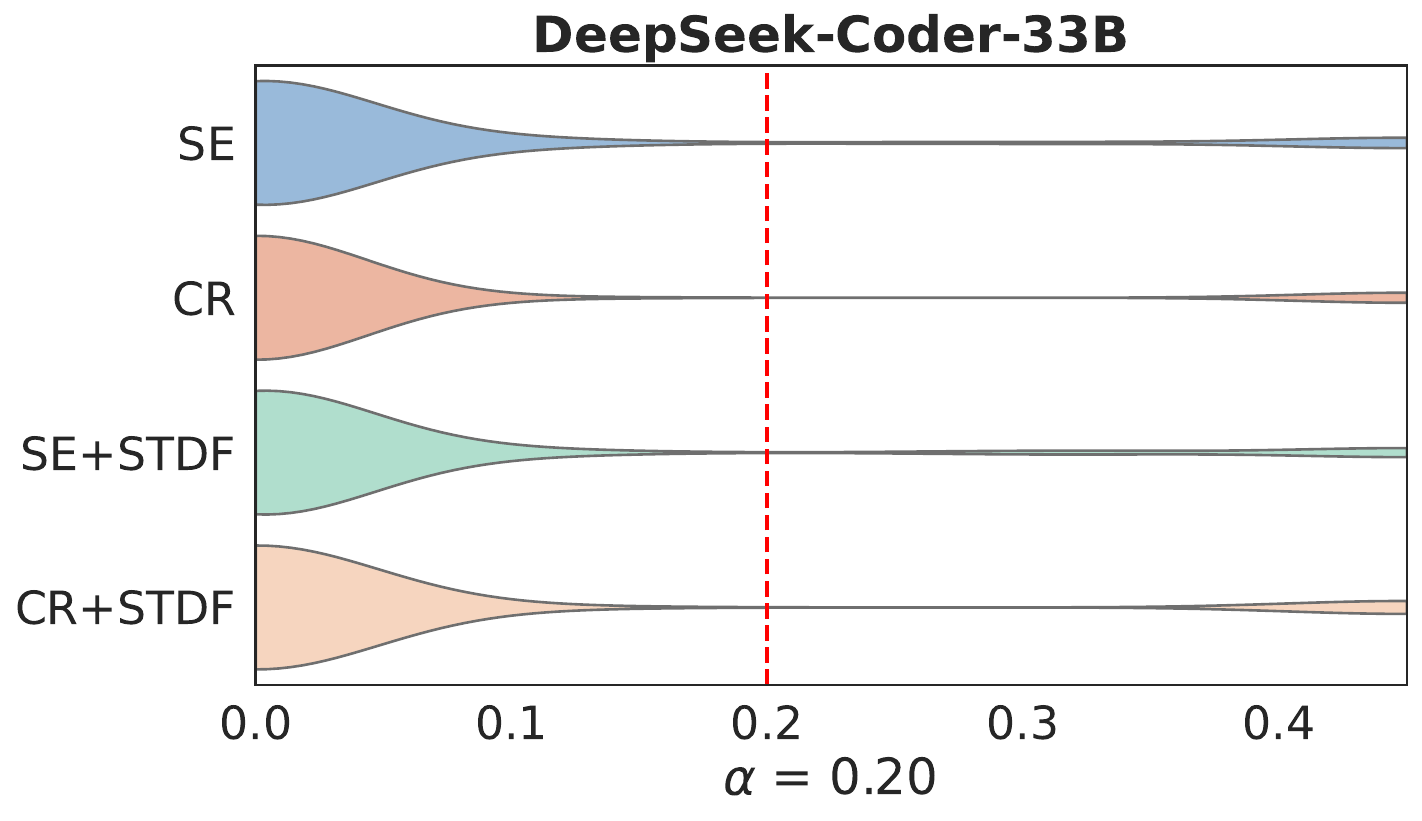}
        \label{fig:third}
    \end{subfigure}
    \hfill
    \begin{subfigure}[b]{0.24\textwidth}
        \centering
        \includegraphics[width=\textwidth]{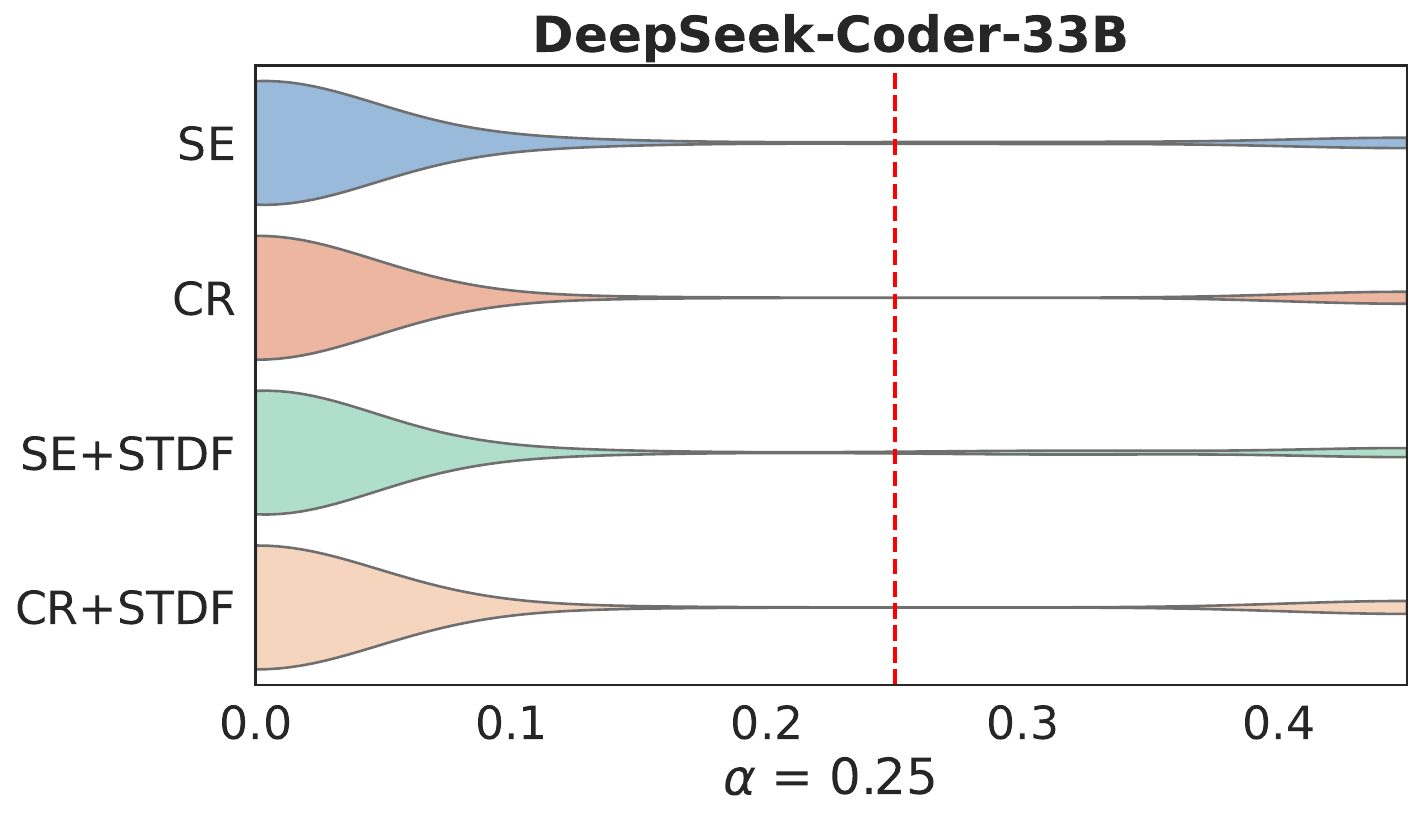}
        \label{fig:fourth}
    \end{subfigure}

    \vspace{0.5em} 

    \begin{subfigure}[b]{0.24\textwidth}
        \centering
        \includegraphics[width=\textwidth]{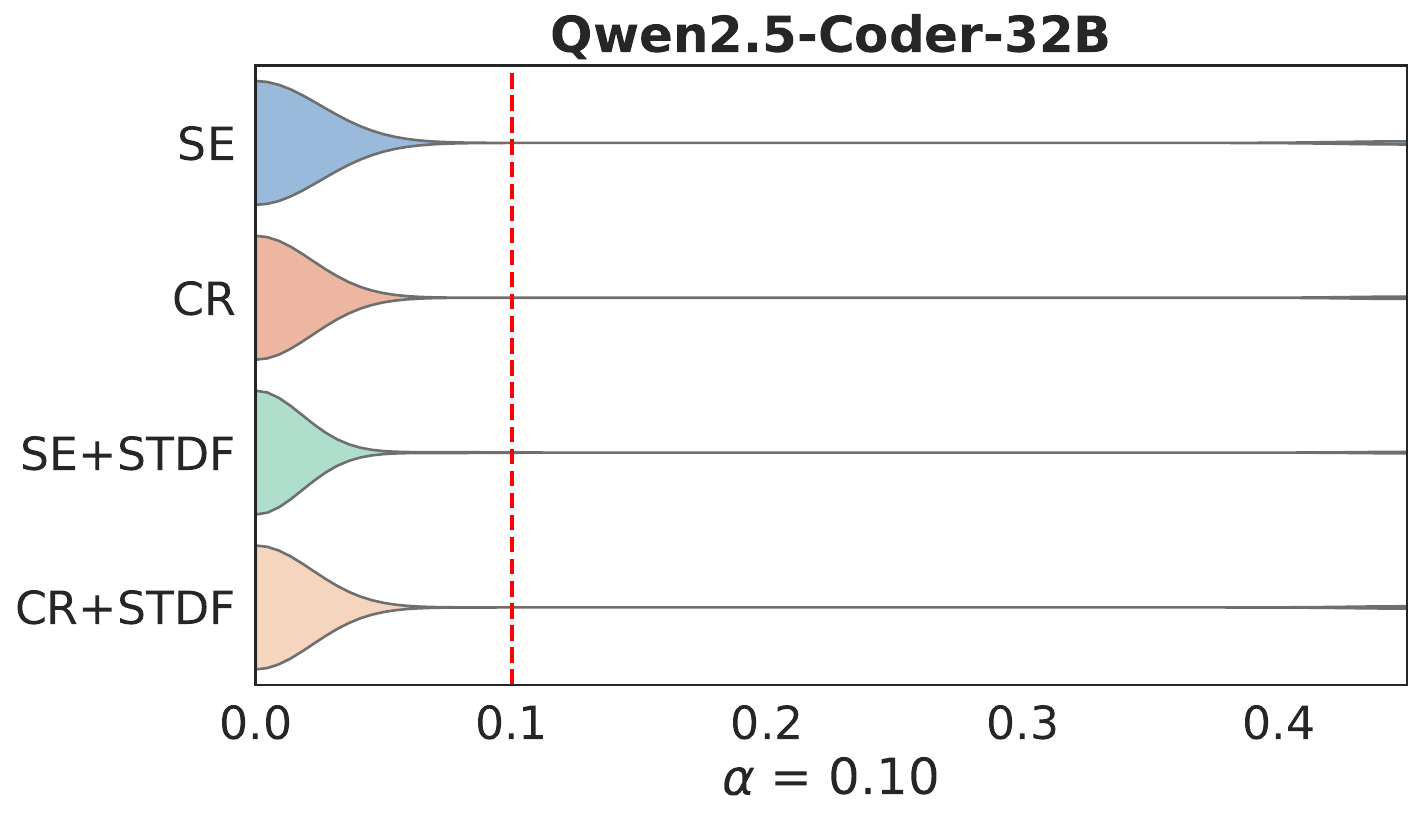}
        \label{fig:fifth}
    \end{subfigure}
    \hfill
    \begin{subfigure}[b]{0.24\textwidth}
        \centering
        \includegraphics[width=\textwidth]{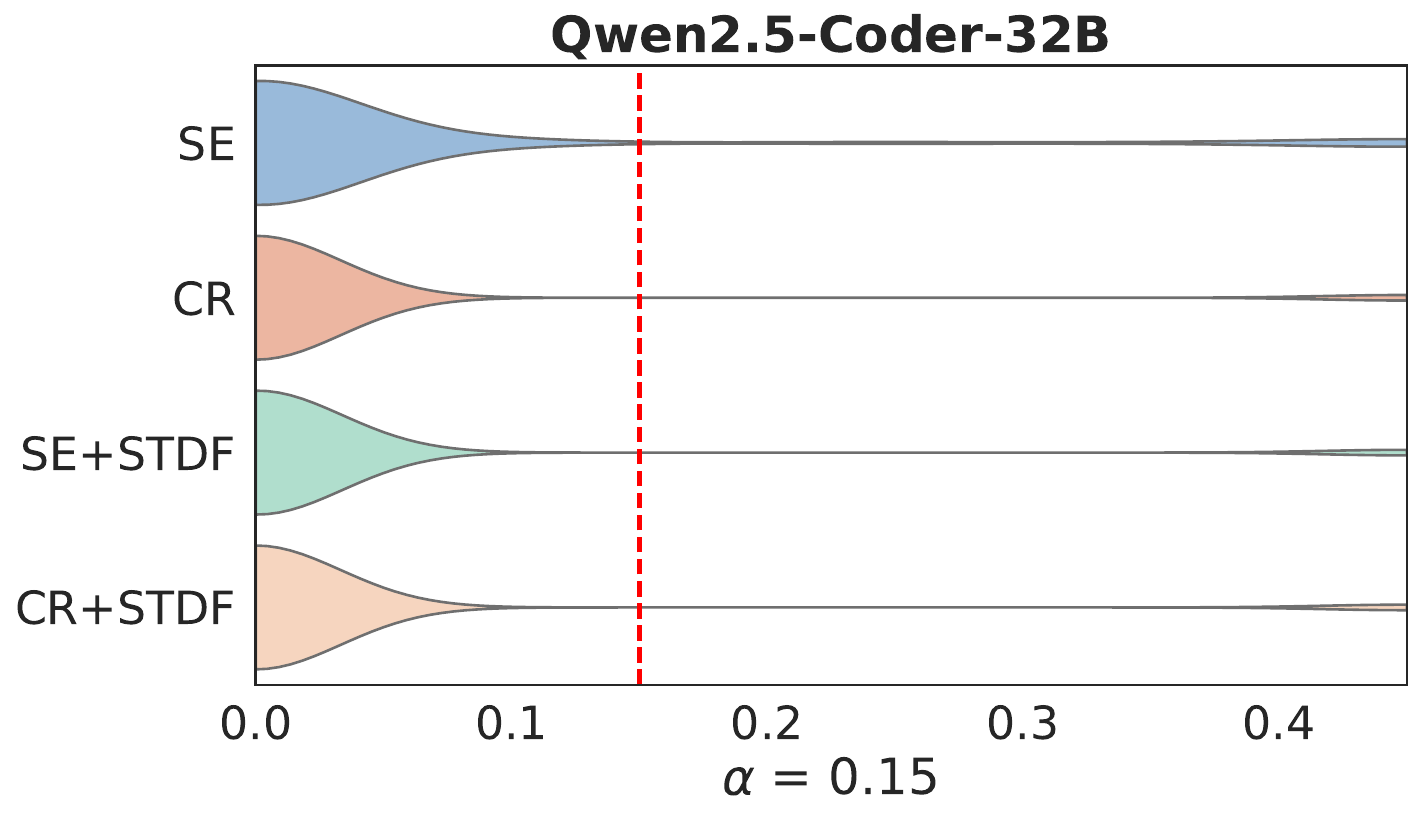}
        \label{fig:sixth}
    \end{subfigure}
    \hfill
    \begin{subfigure}[b]{0.24\textwidth}
        \centering
        \includegraphics[width=\textwidth]{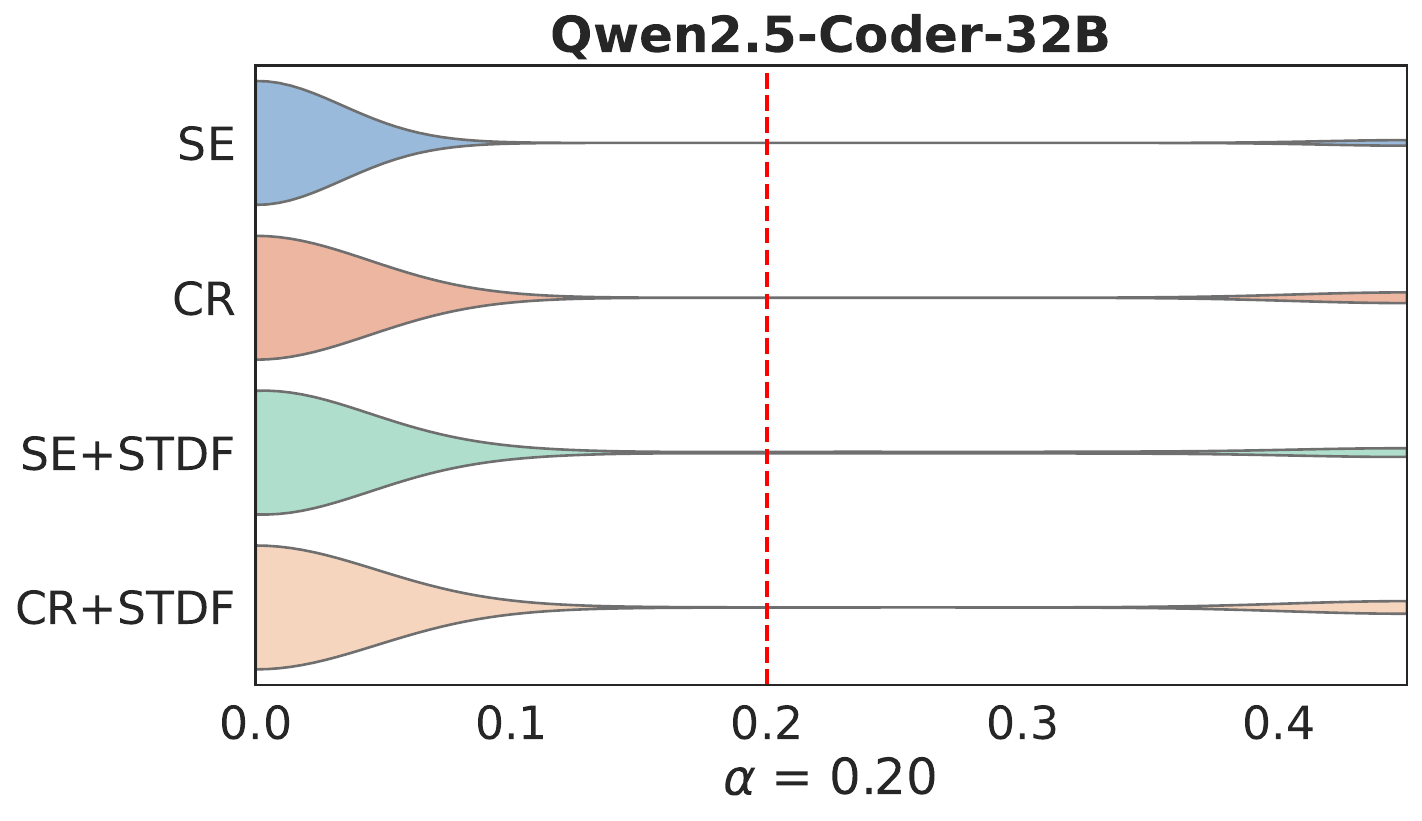}
        \label{fig:seventh}
    \end{subfigure}
    \hfill
    \begin{subfigure}[b]{0.24\textwidth}
        \centering
        \includegraphics[width=\textwidth]{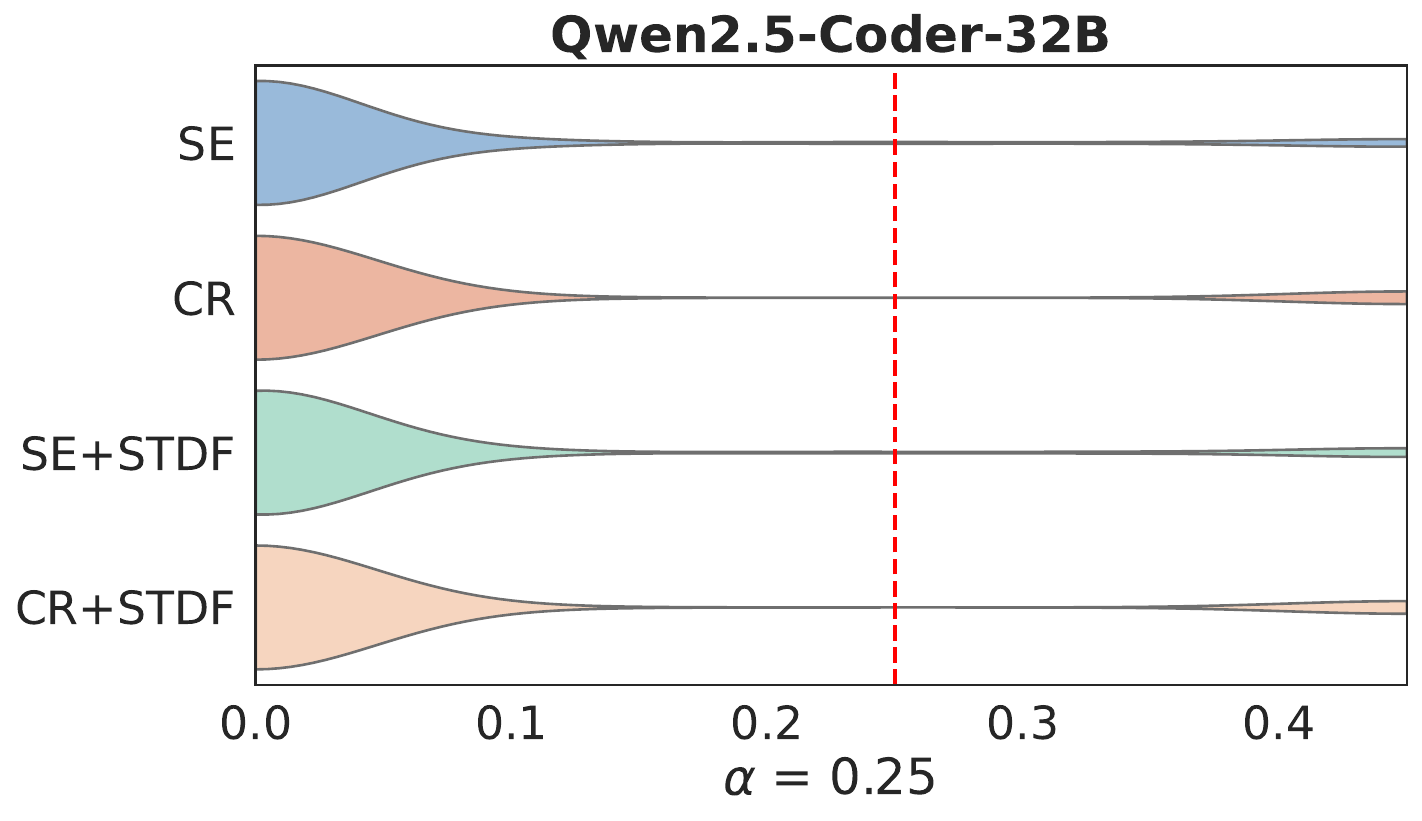}
        \label{fig:eighth}
    \end{subfigure}
    \vspace{-2mm}
    \caption{Admission risk distribution on HumanEval under different risk tolerance $\alpha$.  
    The thickness of each plot corresponds to the density of tasks at that risk level. Most of the admission risks are controlled under the given tolerance (the red dashed line).}
    \label{fig:riskiid}
\end{figure*}

\subsection{Effectiveness Results}
For the score function computation in both the calibration and testing phases, we generate 64 code samples and 64 test cases for each problem.
The results on the HumanEval and MBPP benchmarks are presented in Table~\ref{tab:abstention_main}, which are the average results of three independent runs. 
In the table, `SE' stands for \name with semantic entropy, `CR' stands for \name with cluster ratio, and `SE+STDF' and `CR+STDF' stand for the version when the STDF mechanism is included. 




\smallskip
\noindent {\em (1) The quality of generated tests matters.} The poor performance of both \textit{Execution} and \textit{CodeHalu} demonstrates that LLM-generated tests cannot reliably replace oracle test suites. \textit{Execution} fails to accurately estimate the pass rate ($H@k$), while \textit{CodeHalu}'s performance degrades significantly when forced to treat potentially flawed generated tests as ground truth. These results confirm a critical quality gap, indicating that naive reliance on raw generated tests without rigorous filtering, is insufficient for effective abstention.
In contrast, when the STDF mechanism is integrated, \name consistently yields the best performance across benchmarks. 
Compared with the best existing results, \name achieves an average of 26.5\% and 11.9\% absolute improvements w.r.t. abstention precision and F1-score, respectively.

\smallskip
\noindent {\em (2) Static score functions are less effective in code generation tasks.} 
Among the baselines, \textit{PPL}, \textit{CLM}, and \textit{NLI} are static methods proposed for natural language generation, without actually running the code. The experimental result shows that these methods consistently underperform compared to our execution-based methods (the four bottom rows in the table). 
This confirms that traditional NLP metrics fail to capture the semantic correctness of code, reinforcing the necessity of execution for reliable risk estimation.

One might argue that execution-based approaches inherently incur higher computational costs than static baselines. To demonstrate its necessity, we conducted an experiment where we significantly increased the sampling budget for the static baselines. Specifically, we allowed PPL and CLM to generate $N=256$ samples, while keeping \name (CR+STDF) with the original $N=64$ samples. The results are summarized in Table~\ref{tab:compute_enhanced}, where we use HumanEval and three code LLMs for simplicity. The result reveals that even with quadrupled samples, the performance of static methods remains significantly lower than \name. This confirms that the limitations of static methods stem from their inability to capture semantic correctness, a fundamental deficit that cannot be overcome simply by scaling up the sample size.


\smallskip
\noindent{\em (3) \name transfers effectively across different datasets.}
We also show in Table~\ref{tab:abstention_main} the results
when the calibration is conducted on MBPP and the testing is conducted on HumanEval. Despite the distribution shift, \name maintains robust performance, consistently outperforming baselines even when they are evaluated in-distribution (compared with results of `HumanEval'). This demonstrates that the calibration results of \name exhibit transferability across different datasets.

\subsection{Risk Control Guarantee}
\begin{wrapfigure}{r}{0.48\textwidth}
    \vspace{-6pt}
    \centering
    \begin{minipage}[t]{0.48\linewidth}
        \centering
        \includegraphics[width=\linewidth,height=2.85cm,keepaspectratio]{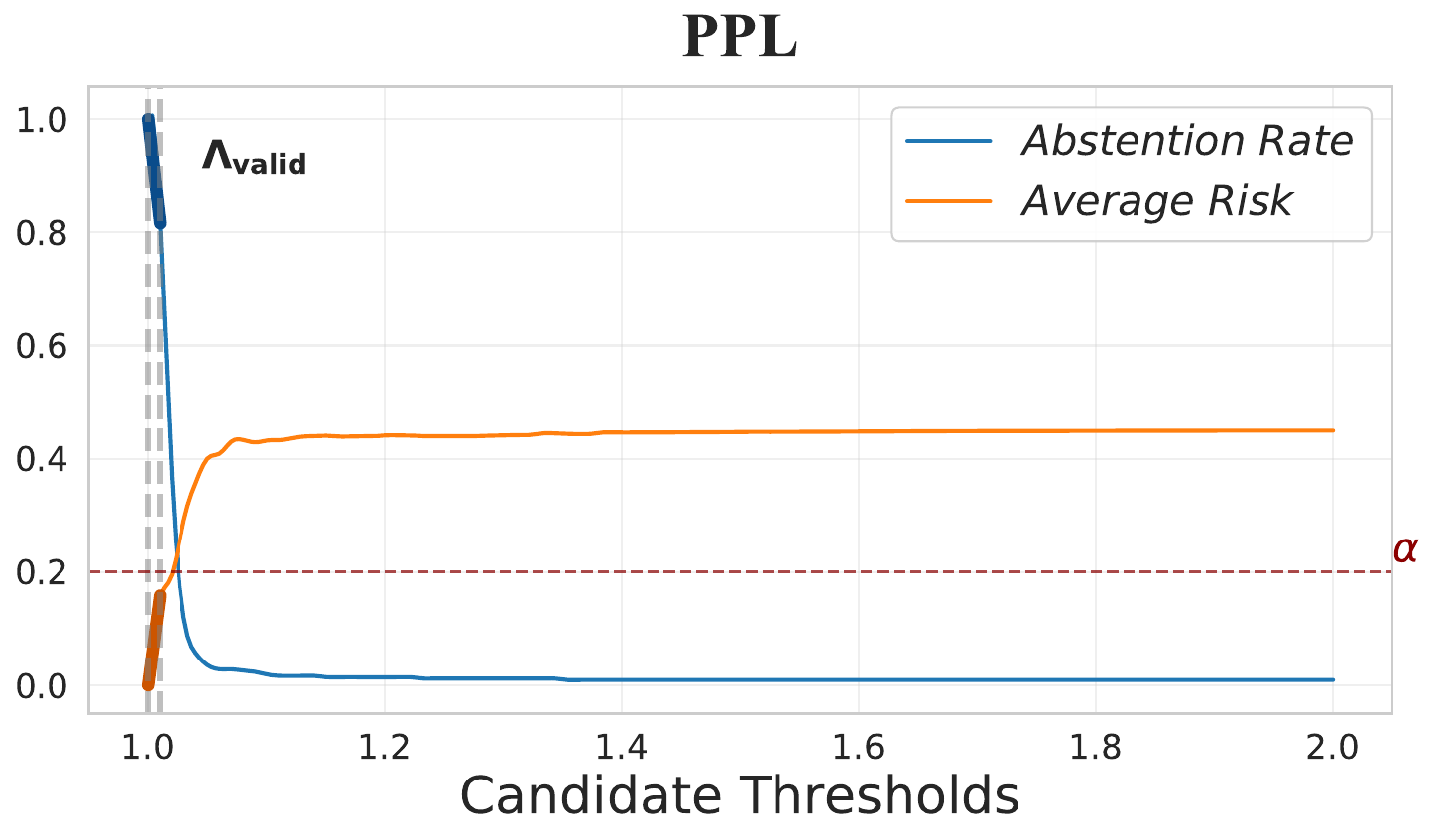}\\[-2pt]
        {\scriptsize PPL}
    \end{minipage}\hfill
    \begin{minipage}[t]{0.48\linewidth}
        \centering
        \includegraphics[width=\linewidth,height=2.85cm,keepaspectratio]{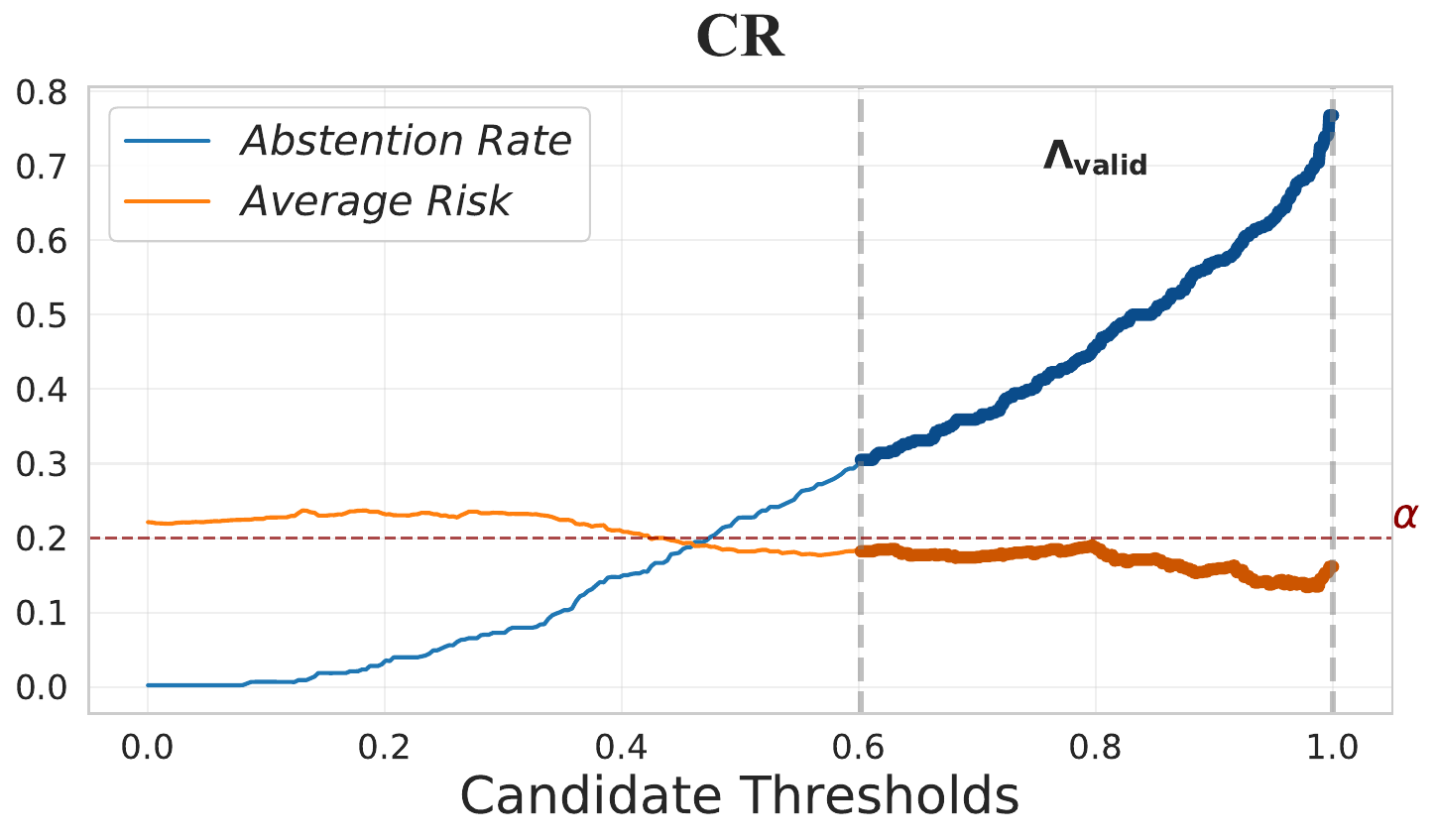}\\[-2pt]
        {\scriptsize CR}
    \end{minipage}
    \caption{Trade-off between abstention rate and admission risk on MBPP (Qwen2.5-Coder-32B). Both methods control risk within the valid range (thick lines); PPL requires a much higher abstention rate (80\% vs.\ 30\%).}
    \label{fig:ar&risk}
    \vspace{-8pt}
\end{wrapfigure}
We next verify the theoretical risk control guarantee. We focus on the in-distribution setting (calibrated and tested on HumanEval) using DeepSeek-Coder and Qwen2.5-Coder. For each tolerance level $\alpha$ (with fixed $\delta=0.1, k=3$), we select the valid threshold $\lambda \in \Lambda_{valid}$ that yields the lowest abstention rate. Fig.~\ref{fig:riskiid} illustrates the distribution of the resulting admission risks. The results confirm that the empirical risk is reliably controlled below the target $\alpha$. Similar results are observed in the transferability setting (calibrated on MBPP and tested on HumanEval), as detailed in Appendix~\ref{appd:transferability}. In summary, \name successfully meets the theoretical guarantees. It remains effective in over 90\% cases in terms of keeping the risk below the desired tolerance level (the area on the left side of the red tolerance line accounts for more than 90\% of the total area).  

Note that one method may simply refuse most code generation tasks to keep the admission risk low. Here,
we also analyze $\Lambda_{valid}$ to understand the trade-off between admission risk and abstention rate. We set the tolerance $\alpha = 0.2$, and compare the results of \name with cluster ratio\footnote{We choose `CR' in this experiment for interpretation, as it has only one parameter.} (i.e., `CR') and {\em PPL}.  The results of Qwen-Coder-32B on the MBPP dataset are shown in Fig.~\ref{fig:ar&risk}.
It can be observed that the admission risk is under control for both methods.
However, {\em PPL} must refuse a large number of tasks (over 80\%) to meet the risk tolerance; in contrast, our method only needs to refuse 30\% of tasks. 

%% file: tables/all_result.tex
\begin{table*}[!t]
\centering
\small 
\caption{Task abstention results on HumanEval and MBPP. Our approaches (i.e., `SE+STDF' and `CR+STDF') generally outperform the competitors. Similarly, when the calibration is conducted on MBPP and the testing is conducted on HumanEval (`MBPP$\rightarrow$HumanEval'), our approaches still outperform the competitors and even achieve close performance to the in-distribution case in most cases.}
\label{tab:abstention_main}
\begin{tabularx}{\textwidth}{l | *{8}{>{\centering\arraybackslash}X}} 
\toprule
\multirow{2}{*}{\diagbox[width=6em, trim=l]{Method}{Model}} 
 & \multicolumn{2}{c}{DeepSeek-Coder} & \multicolumn{2}{c}{Qwen2.5-Coder} & \multicolumn{2}{c}{CodeLlama} & \multicolumn{2}{c}{WizardCoder} \\
 & \textbf{P} & \textbf{F1} & \textbf{P} & \textbf{F1} & \textbf{P} & \textbf{F1} & \textbf{P} & \textbf{F1} \\
\midrule

\multicolumn{9}{c}{\textbf{HumanEval}} \\
\midrule
Execution   & 39.28 & 55.27 & 24.29 & 38.23 & 72.15 & 83.82 & 35.46 & 51.55 \\
CodeHalu    & 38.56 & 55.66 & 22.76 & 36.84 & 73.10 & 82.88 &
33.11 & 49.01 \\
PPL         & 36.41 & 53.39 & 22.22 & 21.43 & 72.29 & 81.67 & 40.90 & 37.11 \\
NLI         & 49.25 & 52.38 & 17.68 & 30.05 & 77.92 & 62.82 & 31.70 & 48.15 \\
CLM         & 36.41 & 53.39 & 22.22 & 21.42 & 71.27 & 64.42 & 37.71 & 51.49 \\
SE          & 61.19 & 65.07 & 38.23 & 41.67 & 85.95 & 88.51 & 52.63 & 62.01 \\
CR          & 60.52 & 68.14 & 44.44 & 49.23 & 89.47 & 89.47 & 47.99 & 56.00 \\
SE + STDF   & 63.63 & 66.67 & 47.36 & 50.91 & \SOTA{91.67} & 89.19 & \SOTA{62.71} & \SOTA{66.67} \\
CR + STDF   & \SOTA{72.00} & \SOTA{69.92} & \SOTA{73.33} & \SOTA{53.85} & \SOTA{91.67} & \SOTA{91.70} & 61.40 & 66.14 \\

\midrule

\multicolumn{9}{c}{\textbf{MBPP}} \\
\midrule
Execution   & 41.31 & 57.42 & 31.50 & 46.11 & 58.35 & 73.01 & 36.08 & 51.75 \\
CodeHalu    & 40.48 & 57.08 & 30.21 & 45.32 & 57.33 & 72.60 & 
32.70 & 48.60 \\
PPL         & 37.16 & 51.11 & 39.28 & 34.55 & 71.24 & 55.47 & 38.46 & 34.33 \\
NLI         & 39.40 & 49.88 & 30.50 & 39.61 & 67.95 & 58.29 & 33.13 & 36.98 \\
CLM         & 36.32 & 53.28 & 31.60 & 44.25 & 56.60 & 72.28 & 35.10 & 50.00 \\
SE          & 64.70 & 64.61 & 40.74 & 45.65 & 71.77 & 79.02 & 60.16 & 61.26 \\
CR          & 61.99 & 66.28 & 40.00 & 45.58 & 67.24 & 65.67 & 62.99 & 62.50 \\
SE + STDF   & \SOTA{72.65} & \SOTA{72.88} & 48.41 & \SOTA{52.84} & 72.70 & 76.69 & 60.26 & \SOTA{64.70} \\
CR + STDF   & 66.43 & 66.48 & \SOTA{48.94} & 47.44 & \SOTA{79.40} & \SOTA{79.69} & \SOTA{69.23} & 63.07 \\
\midrule

\multicolumn{9}{c}{\textbf{MBPP} $\rightarrow$ \textbf{HumanEval}} \\
\midrule
 SE   &  60.27  &  62.11  & 32.65   &  41.50  &  76.03 &  78.63  &  60.00  &  58.99 \\
 CR   &  63.01  &  65.08  &  32.65  &  37.68  &  79.79  &  79.48  &  57.41  & 56.88 \\
 SE + STDF   &  \SOTA{71.11}  & \SOTA{70.00} &  39.39  &  \SOTA{50.00}  &  80.28  &  77.27  &  68.42  & 59.65 \\
 CR + STDF   &  66.67  &  66.67  & \SOTA{40.63}  & 40.00   &  \SOTA{91.51}  &  \SOTA{91.23}  &  \SOTA{72.10}  & \SOTA{62.39} \\

\bottomrule
\end{tabularx}
\end{table*}

%% file: tables/static_compare.tex
\begin{table*}[htbp]
\centering
\small
\caption{Performance comparison when significant more samples are used for static methods ($N=256$ vs. $N=64$). Static methods fail to narrow the performance gap even with $4\times$ sample budget.}
\label{tab:compute_enhanced}
\begin{tabularx}{\textwidth}{l | *{8}{>{\centering\arraybackslash}X}}
\toprule
\multirow{2}{*}{{Method}} & \multicolumn{2}{c}{{DeepSeek-Coder}} & \multicolumn{2}{c}{{Qwen2.5-Coder}} & \multicolumn{2}{c}{{WizardCoder}} \\
& \textbf{P} & \textbf{F1} & \textbf{P} & \textbf{F1} & \textbf{P} & \textbf{F1} \\
\midrule
PPL ($N=256$) & 37.02 & 54.36 & 17.79 & 30.21 & 39.42 & 52.22 \\
CLM ($N=256$) & 36.41 & 53.39 & 20.32 & 32.89 & 36.29 & 51.64 \\
\rowcolor{gray!10} \textbf{\name} ($N=64$) & \textbf{72.00} & \textbf{69.92} & \textbf{73.33} & \textbf{53.85} & \textbf{61.40} & \textbf{66.14} \\
\bottomrule
\end{tabularx}
\end{table*}

%% file: src/conclusion.tex
\section{Conclusion} \label{sect:conc}
In this paper, we introduce and study the task abstention problem for LLM-based code generation, i.e., determining whether a given LLM should abstain from performing a specific code generation task to avoid potential hallucination. 
Our approach 
features a calibrated abstention rule, grounded in the principles of multiple hypothesis testing. 
A distinguished advantage is that it provides a rigorous, distribution-free theoretical guarantee on its abstention decisions, whose effectiveness is also confirmed by the experiments. Our work represents progress in the pursuit of \emph{provably correct} LLM code generation. 

%% file: src/related_work.tex
\section{Related Work} \label{appd:related}

\noindent\textbf{Code Hallucination.}
The phenomenon of code hallucination, where LLMs generate code that is illogical, incorrect, or unfaithful to user requirements~\cite{Fan2023LLM4se}, presents a challenge to ensure the accuracy, reliability and security of AI-generated code~\cite{agarwal2024codemirage, eghbali2024hallucinator, tian2025codehalu}. Existing research has explored code hallucination from several angles, often categorizing failures based on when they occur. These categories primarily include \emph{syntactic hallucination}, which are errors that violate the programming language's syntax and prevent code from being compiled or interpreted~\cite{agrawal2023monitorguided, Fan2023autorepair, wang2025understandingcharacteristicscodegeneration}; \emph{runtime hallucination}, where the code is syntactically valid but produces errors such as exceptions or crashes~\cite{Fan2023autorepair, liu2024exploring, tian2025codehalu, zhang2025llmhalu}, and \emph{functional hallucination}, where code that runs without error fails to meet the program's intended requirements~\cite{Fan2023autorepair, liu2024exploring, wang2025understandingcharacteristicscodegeneration, zhang2025llmhalu}, among others. 

To facilitate a more rigorous evaluation of hallucination in LLM-based code generation, a number of benchmarks have been developed. Notable examples include CodeHaluEval~\cite{tian2025codehalu}, CodeMirage~\cite{agarwal2024codemirage}, MultiPL-E~\cite{Cas2023multipl-E}, HalluCode~\cite{liu2024exploring}, etc. Prior work has also proposed various strategies to mitigate code hallucination. The De-hallucinator~\cite{eghbali2024hallucinator} pre-indexes a project's codebase and uses Retrieval-Augmented Generation (RAG) to inject relevant APIs into prompts; Liu et al.~\cite{liu2024refine} leverage the LLM's self-revision capabilities by providing it with feedback based on static analysis; SynCode~\cite{ugare2025syncode} uses a formal grammar representation (EBNF) to guide the model's decoding and ensure syntactic validity; 
ClarifyGPT~\cite{Mu2024clarifygpt} introduced a framework where the LLM proactively asks clarifying questions to help users refine their initial prompts. Different from the above work that focuses on the sample-level hallucination problem, we study the task abstention problem.

\noindent\textbf{LLM Abstention.}
Abstention 
is increasingly recognized for its potential to mitigate hallucination and enhance safety in LLM systems~\cite{wen-etal-2025-know, varshney-etal-2024-art, wang-etal-2024-answer, zhang-etal-2024-r}. A guiding principle is that a system should abstain when it is insufficiently confident in the correctness of its output or if there is a high probability of error~\cite{ahdr2024distinguish, kim-thorne-2024-epistemology, cao-2024-learn}. Existing work has proposed various strategies to determine when the LLM should abstain, targeting different stages of the model lifecycle. 

During the training and alignment phase, Neeman et al.~\cite{neeman-etal-2023-disentqa} use data augmentation to fine-tune models to recognize unanswerable questions. Yang et al.~\cite{yang2024alignment} construct ``honesty'' alignment datasets by substituting a model's incorrect response with ``I don't know'' and then fine-tuning on this revised data.
At inference time, a common approach is to use post-processing techniques based on model uncertainty. These include calculating the log probability of a `True' token via indirect logit methods~\cite{lin2022teaching, tian2023just}, using a surrogate LLM to approximate the confidence of a black-box model~\cite{shrivastava2024llamas}, or assessing the semantic entropy of responses~\cite{kuhn2023semantic}. A different inference-time strategy involves LLM collaboration, where a second ``test'' LLM is employed to examine the output of the first, helping identify harmful queries or correct the initial response before it is shown to the users~\cite{feng-etal-2024-dont, chen2023jailbreaker}.
The existing methods are mainly proposed for the natural language generation problem. In this work, we argue that test code generation and execution are essential for the code generation task abstention problem.

\smallskip
\noindent\textbf{Conformal Risk Control.}
Approaches based on Conformal Prediction (CP) are known for their ability to select a guaranteed threshold by analyzing the Guantile distribution of risk term on a calibration set~\cite{2019conformalshift, Rina2023Conformal, gibbs2021adaptive}. However, standard CP methods are often constrained to operating on a single candidate threshold at a time. They are not equipped to handle a \emph{vector} of thresholds derived simultaneously from multiple score functions, which limits their applicability in scenarios requiring multifaceted evaluation. 

A related framework, Conformal Risk Control (CRC), also efficiently utilizes calibration sets to manage the prediction risk~\cite{AngelopoulosBFL24}. Its primary limitation, however, is the requirement of a monotonic relationship between the risk function and the threshold. This monotonicity assumption frequently does not hold for the intricate score functions used in complex generative tasks, posing a significant challenge to adapting CRC to our domain.

\smallskip
\noindent\textbf{Confidence Estimation for LLMs.}
Confidence estimation for LLMs aims to provide a measure of predictive uncertainty for LLM outputs. Well-calibrated confidence
can further contribute to migrating the bias and alleviating the hallucination~\cite{geng2023survey, zheng2023large, bubeck2023sparks}.
Recent research on LLM confidence estimation broadly falls into the following veins.
{\em Perplexity confidence}~\cite{huang2023look, duan2023shifting} derives confidence from the probabilities assigned to generated tokens, employing the geometric mean (i.e., perplexity~\cite{chen1998evaluation, blei03LDA}) to mitigate sensitivity to output length; {\em verbalized confidence}~\cite{kadavath2022language, xiong2023can, tian2023just} directly prompts the LLM to explicitly express its confidence alongside its answer (e.g., ``Read the question and give your answer and corresponding confidence score''); {\em self-consistency confidence}~\cite{xiong2023can, yadkori2024believe, becker2024cycles}
assesses confidence by having the LLM generate multiple answers for the same input and then 
measuring the consistency among them~\cite{wang2022self, chen2023universal, cheng2024relic}, with higher consistency indicating greater confidence.




%% file: src/LTT.tex
\section{The LTT Framework}\label{appd:ltt}
The Learn Then Test (LTT) framework~\cite{angelopoulos2021learn} is designed to provide statistical guarantees for machine learning models by simply adding a post-processing step on a calibration set after the model is trained.
Consider the task where each instance $x \in \mathcal{X}$ is associated with a ground-truth label $y \in \mathcal{Y}$.
Let $\mathcal{D}_{cal}=\{(x_i, y_i)\}_{i=1}^m \subseteq \mathcal{X} \times \mathcal{Y}$ be a calibration set composed of the input $x$ and its ground-truth label $y$, which are independently and identically distributed (i.i.d.) drawn. A post-processing function $\mathcal{T}_{\lambda}: \mathcal{X} \to \mathcal{Y}'$ with parameter $\lambda$ is designed to map $\mathcal{X}$ to any space $\mathcal{Y}'$. (For instance, in classification $\mathcal{Y}'$ may be defined as $2^\mathcal{Y}$, i.e., all possible subsets of $\mathcal{Y}$.) In other words, instead of predicting a label for each instance, LTT aims to predict a subset of labels so that the true label is within the subset with a high probability. The choice of the subsets is decided by $\lambda$ (e.g., labels with predictive probability greater than $\lambda$ are included in the subset). 

Based on the post-processing function $\mathcal{T}_{\lambda}$, a \emph{risk}  $R(\mathcal{T}_{\lambda}(x)) \in \mathbb{R}$ on a given $x$ can be defined to measure the task-specific statistical error (e.g., the miscoverage rate of true labels in $\mathcal{Y}'$ for the classification task). Since the risk is mainly decided by parameter $\lambda$, we rewrite $R(\mathcal{T}_{\lambda}(x))$ as $R(x; \lambda)$ for brevity.

The goal of LTT is to ensure the guarantee as stated in Eq. ~\eqref{eq:LTT}.
Using the classification task as an example, intuitively Eq.~\eqref{eq:LTT} asserts that the risk of a wrong classification (e.g., the true label is not in the output subset) is below the threshold $\alpha$ with probability at most  $\delta$.

The core of LTT is to obtain the set $\Lambda_{valid}$. For this purpose,  we can traverse all the plausible $\lbrace \mathcal{T}_{\lambda}\rbrace_{\lambda \in \Lambda}$ and estimate their risk on the calibration set $\mathcal{D}_{cal}$ using multiple hypothesis testing. Specifically,
for each $\lambda_j$ in a discrete set $\Lambda = \lbrace \lambda_1, \cdots, \lambda_N \rbrace$, we have a null hypothesis $\mathcal{H}_j : R(\lambda_j) > \alpha$ and $\mathcal{H}_j$ is rejected when $\lambda_j$  controls the risk, i.e., $\lambda_j\in \Lambda_{valid}$.  For each null hypothesis, we can compute a finite-sample valid $p$-value using a concentration inequality~\cite{angelopoulos2021learn}. $\Lambda_{valid}$ can then be calculated by applying any \emph{family-wise error rate} (FWER)-controlling algorithm, which receives the set of $p$-values and returns the set of $\lambda$ that we should reject the associated null hypothesis. For example, the Bonferroni correction is a typical FWER-controlling algorithm which yields $\Lambda_{valid} = \left\lbrace \lambda_j: p_j \le \frac{\delta}{|\Lambda|} \right\rbrace$.

Once $\Lambda_{valid}$ is obtained on the calibration set, LTT ensures that the following theorem holds for an i.i.d. test sample $x_{test}$.
\begin{theorem}[Learn Then Test~\cite{angelopoulos2021learn}] \label{th:ltt}
    Suppose $p_j$ is super-uniform for all $j$ under $\mathcal{H}_{j}$, and assume a valid FWER-controlling algorithm at level $\delta$. Then Eq.~\eqref{eq:LTT} holds for the test sample $x_{test}$. 
\end{theorem}

\noindent {\em Key insight of applying LTT.} The key insights of using LTT in our task abstention problem for LLM-based code generation are as follows. LTT was originally proposed to generate a set of responses, instead of a single response, so that the true response is within the response set with a guaranteed probability. In code generation, the LLM can refuse the code generation task if the response set is empty after a few attempts. Meanwhile, the statistical guarantee from LTT still stands. 


%% file: tables/transfer_pic.tex
\begin{figure*}[t]
    \centering
    \begin{subfigure}[b]{0.24\textwidth}
        \centering
        \includegraphics[width=\textwidth]{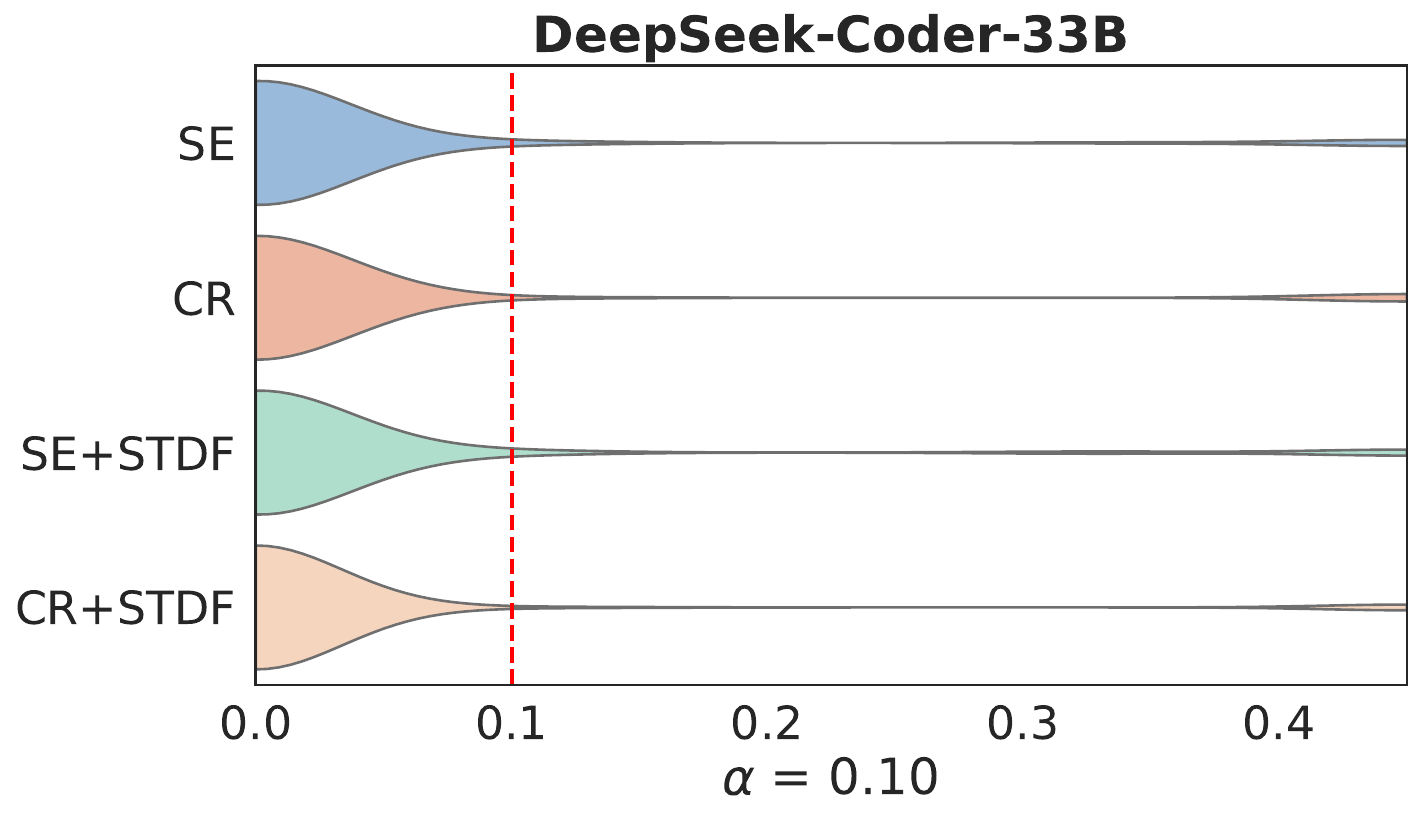}
    \end{subfigure}
    \hfill
    \begin{subfigure}[b]{0.24\textwidth}
        \centering
        \includegraphics[width=\textwidth]{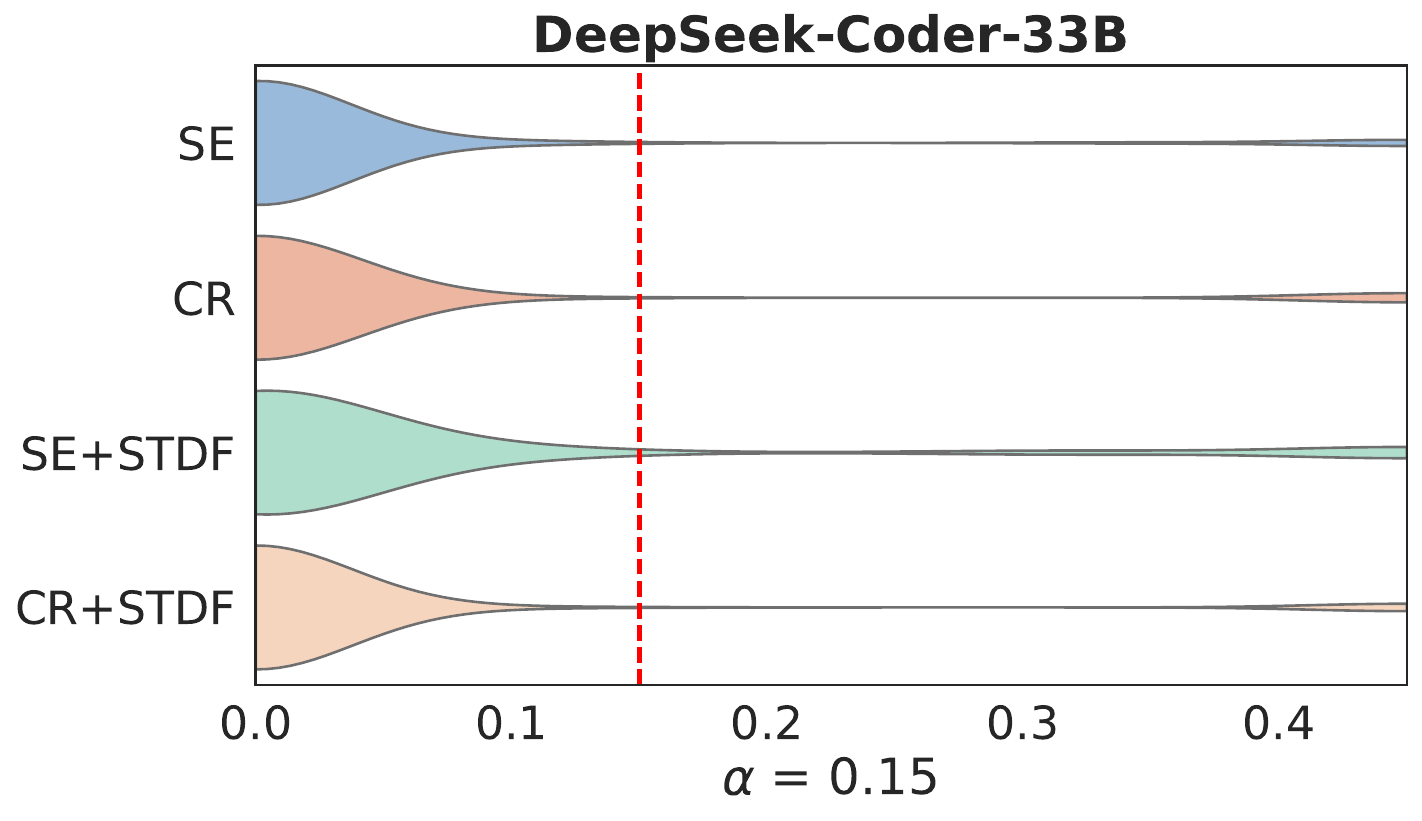}
    \end{subfigure}
    \hfill
    \begin{subfigure}[b]{0.24\textwidth}
        \centering
        \includegraphics[width=\textwidth]{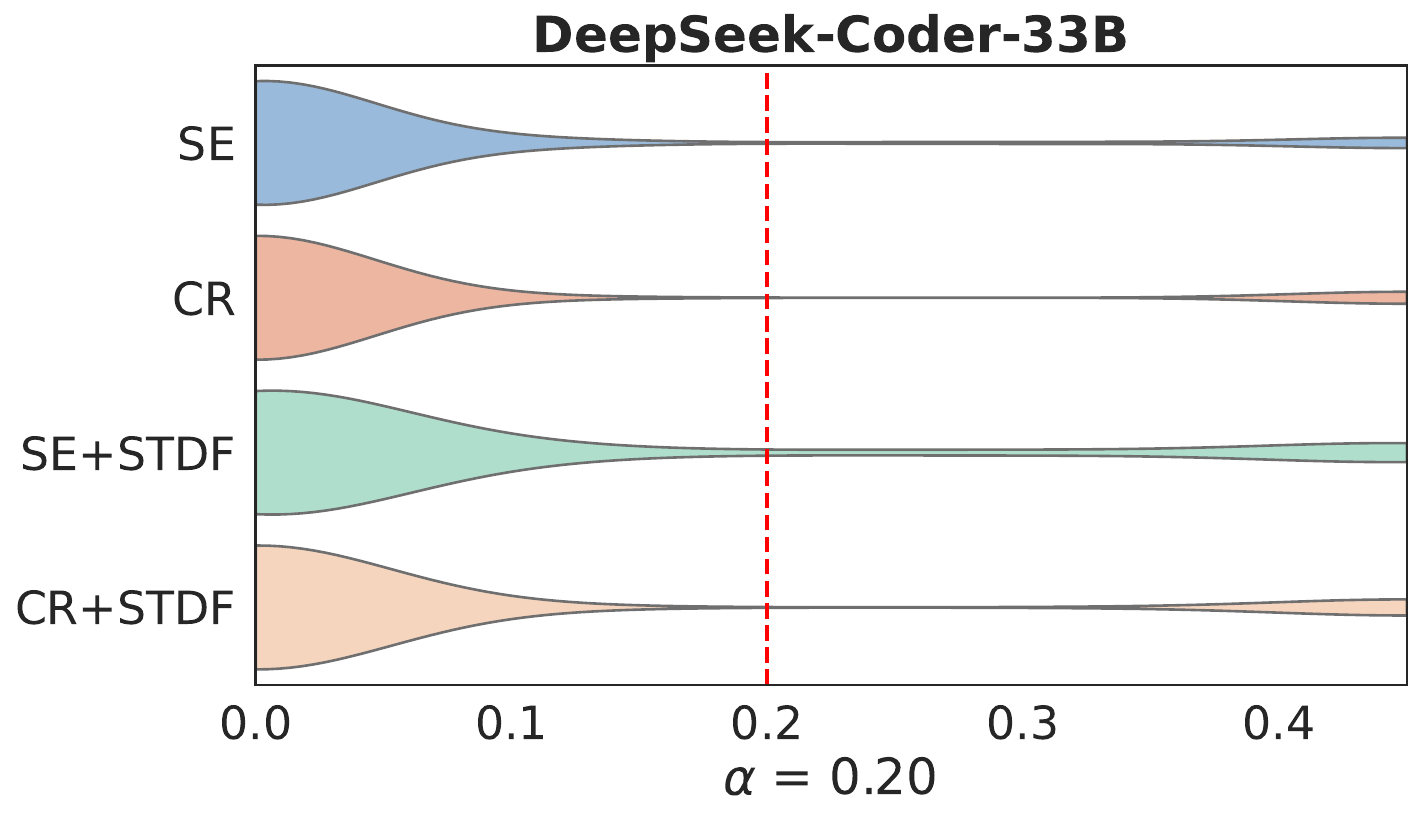}
    \end{subfigure}
    \hfill
    \begin{subfigure}[b]{0.24\textwidth}
        \centering
        \includegraphics[width=\textwidth]{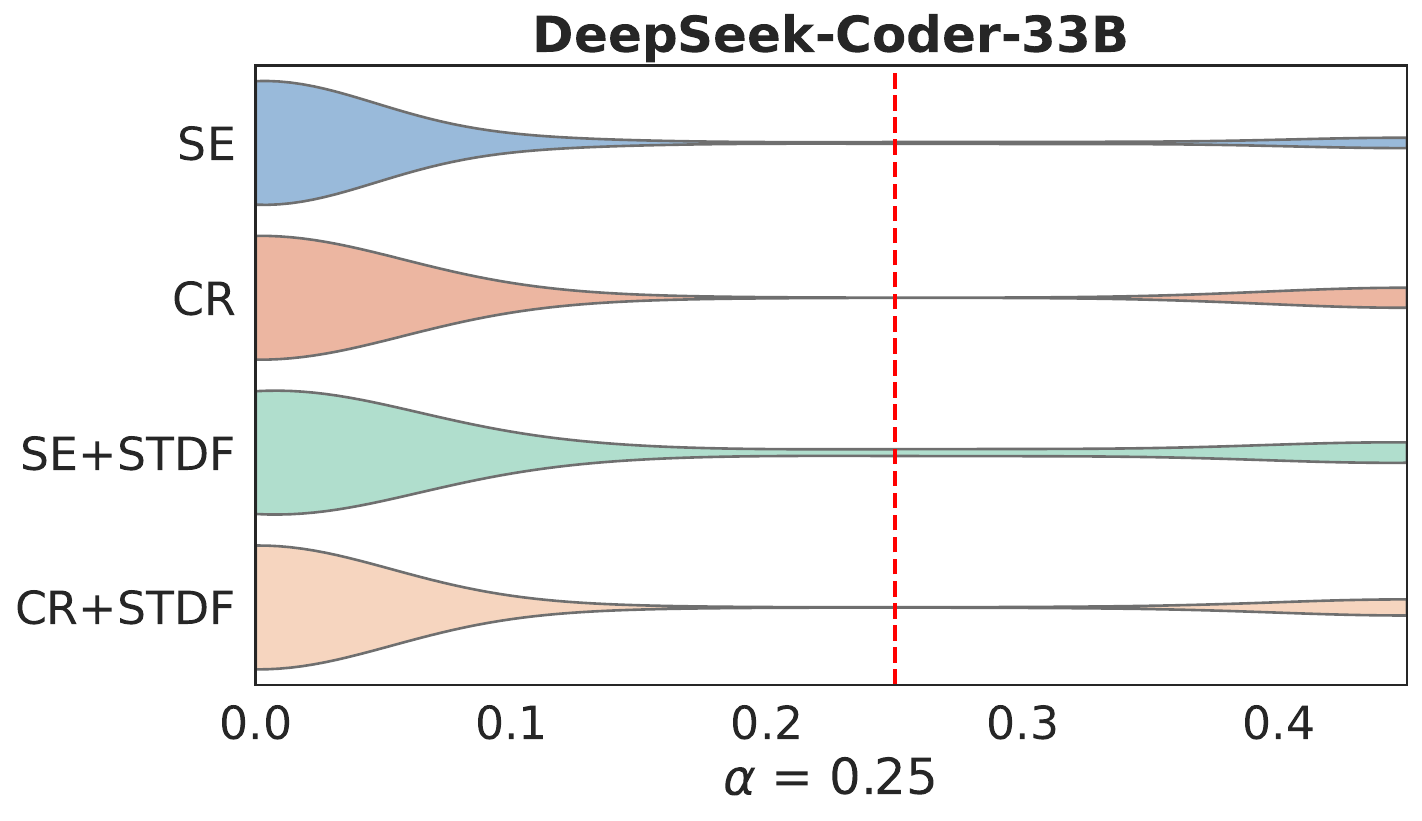}
    \end{subfigure}

    \vspace{0.5em} 

    \begin{subfigure}[b]{0.24\textwidth}
        \centering
        \includegraphics[width=\textwidth]{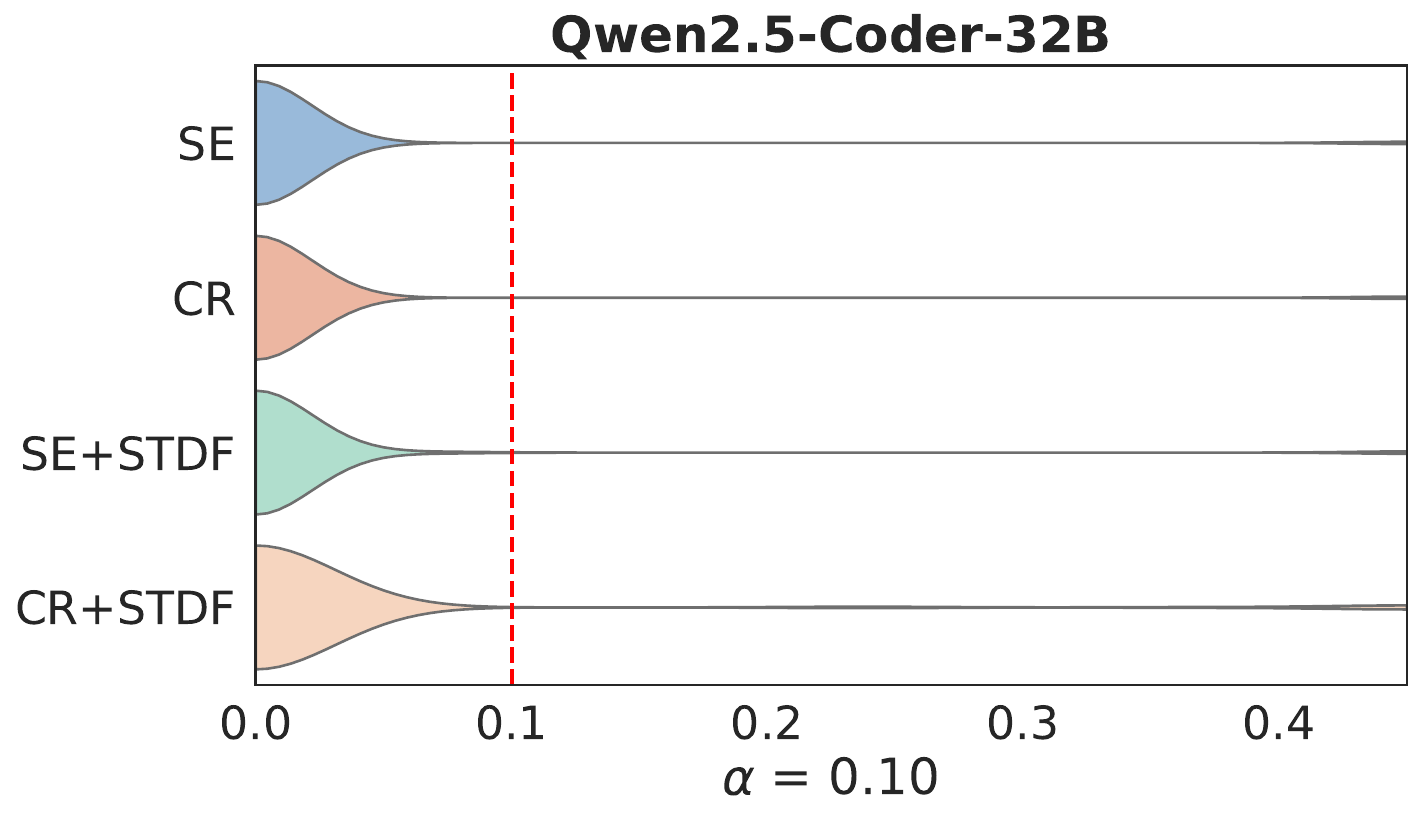}
    \end{subfigure}
    \hfill
    \begin{subfigure}[b]{0.234\textwidth}
        \centering
        \includegraphics[width=\textwidth]{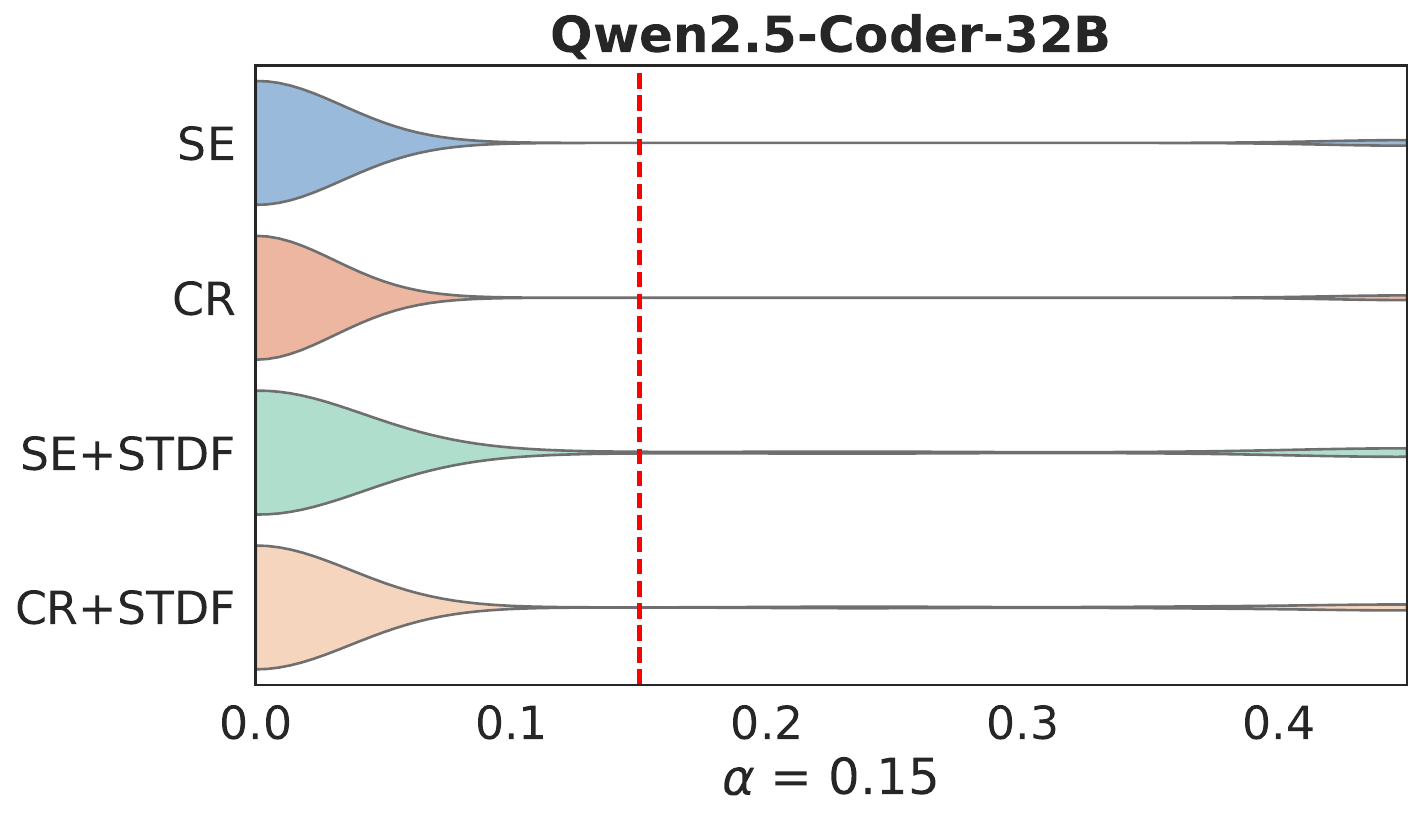}
    \end{subfigure}
    \hfill
    \begin{subfigure}[b]{0.24\textwidth}
        \centering
        \includegraphics[width=\textwidth]{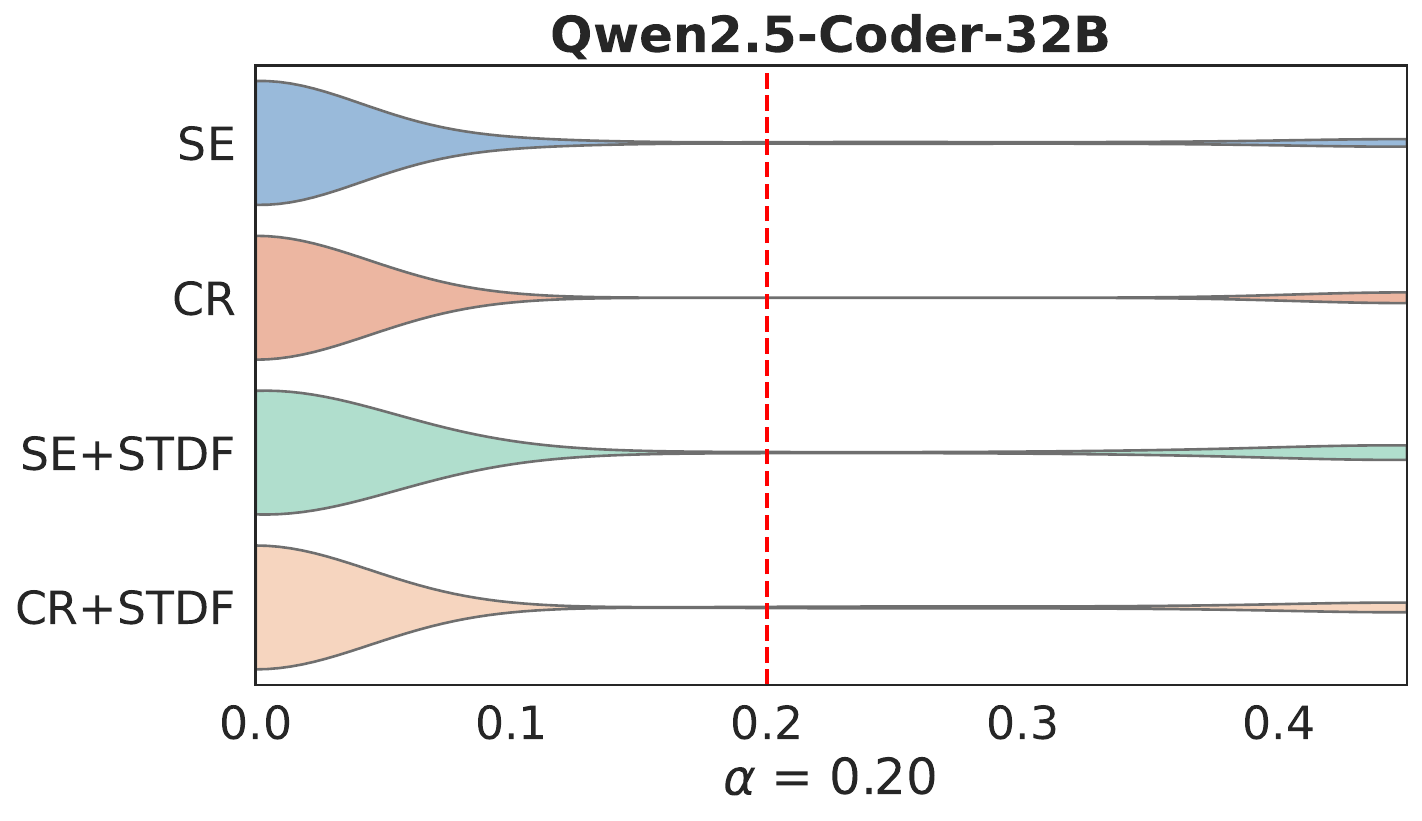}
    \end{subfigure}
    \hfill
    \begin{subfigure}[b]{0.24\textwidth}
        \centering
        \includegraphics[width=\textwidth]{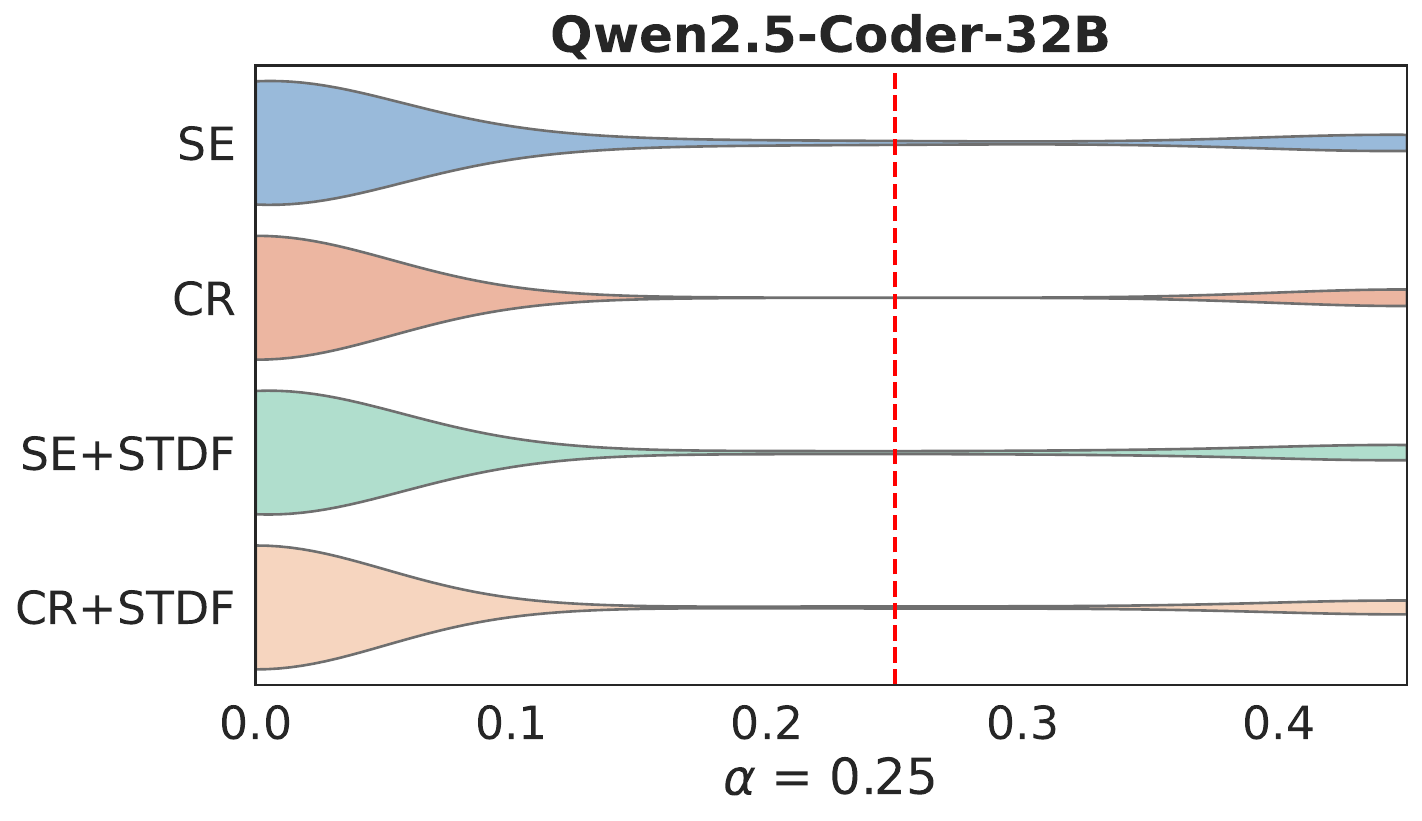}
    \end{subfigure}
    \caption{Admission risk distribution on HumanEval under different risk tolerance $\alpha$, when the LLM is calibrated on MBPP. Most of the admission risks are still under the given tolerance.}
    \label{fig:risknoniid}
\end{figure*}

%% file: src/scoredef.tex
\section{Score Function Definitions} \label{appd:score_functions}
In Section~\ref{sec:3.3}, we introduced two score functions based on execution-based clustering. Here, we provide their formal definitions.
Let $Y=\{y_1, \dots, y_n\}$ be the generated code samples and $\mathcal{C} = \{C_1, \dots, C_h\}$ be the partition of $Y$ derived from execution results on test cases $T$.

\subsection{Confidence-based Score Function}
The consistency function $\mathbb{I}(y_1, y_2; \mathcal{C})$ between two code samples is defined as:
\begin{equation}
    \mathbb{I}(y_1, y_2; \mathcal{C}) = 
    \begin{cases}
    1, & \exists C_j \in \mathcal{C}, \{y_1, y_2\} \subseteq C_j \\
    0, & \text{otherwise}
    \end{cases}
\end{equation}
The confidence score for a sample $y \in Y$ is the average consistency with all other samples:
\begin{equation}
    Conf(y) = \frac{1}{n} \sum_{i=1}^n \mathbb{I}(y, y_i; \mathcal{C}).
\end{equation}
A high confidence score indicates that the sample belongs to a dominant semantic cluster, reflecting the model's conviction in that specific logic~\cite{chen2022codetcodegenerationgenerated}.

\subsection{Semantic Entropy-based Score Function}
Semantic entropy measures the diversity of the semantic clusters. For a task $x$, it is defined as:
\begin{equation}
\begin{split}
SE(x) &= -\sum_{C_i \in \mathcal{C}} p(C_i \mid x) \log{p(C_i \mid x)} \\
      &\approx -|\mathcal{C}|^{-1} \sum_{C_i \in \mathcal{C}} \log{p(C_i \mid x)}.
\end{split}
\end{equation}
A higher entropy value signifies greater inconsistency among the generated solutions, suggesting a higher likelihood of task-level hallucination~\cite{kuhn2023semantic}.

%% file: src/STDF.tex
\section{Implementation Details of STDF} \label{appd:STDFdetail}

In this section, we provide the formal procedure for the Sample-Test Dual Filtering (STDF) mechanism, as outlined in Algorithm~\ref{alg:STDF}.

The algorithm requires three calibrated thresholds: $\lambda_1$ (maximum error rate tolerance), $\lambda_2$ (maximum pruning ratio), and $\lambda_3$ (maximum entropy tolerance). The process consists of two phases:

\noindent \textbf{Phase 1: Error Rate Pruning (Lines 4--6).} 
We first calculate the error rate for each test case $t$ across all generated code samples $Y$. If the proportion of samples that crash or fail on $t$ exceeds $\lambda_1$, the test case is immediately discarded. This step effectively removes "toxic" inputs that cause widespread crashes.

\noindent \textbf{Phase 2: Diversity-based Pruning (Lines 8--16).} 
For the surviving test cases, we compute the semantic entropy (SE) of their execution outputs. A high SE implies that the code samples produce a wide variety of inconsistent outputs for this specific input, suggesting the test case is ambiguous or invalid. 
However, high variance might also stem from the model's own uncertainty rather than the test's flaw. To avoid aggressively discarding valid tests (which would artificially inflate the consensus score), we impose a safety limit: we sort the test cases by their entropy and remove at most $\lfloor \lambda_2 \cdot |T| \rfloor$ test cases, and only if their entropy exceeds $\lambda_3$.

\begin{algorithm}[h]
\caption{\small Sample-Test Dual Filtering Mechanism}\label{alg:STDF}
\begin{algorithmic}[1]
\Require Generated code samples ${Y}$; generated test cases ${T}$; thresholds $\pmb{\lambda} = [\lambda_1, \lambda_2, \lambda_3]$
\Ensure The refined test cases ${T'}$
\Procedure{STDF}{${Y}, {T}, \pmb{\lambda}$}
\State $res \gets \text{Exec}({Y}, {T})$
\State $FilterSet \gets \emptyset$
\For{each test case $t \in {T}$}
    \If{$\text{ErrorRate}(res[t]) > \lambda_1$}
        \State Erase $t$ from ${T}$ 
    \Else
        \State ${C}_t \gets $ Clustering ${Y}$ by $res[t]$
        \State add $\lbrace t, \text{SE}({C}_t)\rbrace$ to $FilterSet$
    \EndIf
\EndFor
\State sort $FilterSet$ in descending order of SE
\State $max_{num} \gets \lfloor \lambda_2 \cdot |{T}| \rfloor$ 
\For{$t, SE_t$ in top $max_{num}$ elements of $FilterSet$}
    \If{$SE_t > \lambda_3$}
        \State Erase $t$ from ${T}$ 
    \EndIf
\EndFor
\State \textbf{return} ${T}$
\EndProcedure    
\end{algorithmic}
\end{algorithm}

%% file: src/extra_eval.tex
\section{Additional Experimental Analysis} \label{appd:additional_analysis}

In this section, we provide further analysis to validate the robustness and efficiency of \name. Specifically, we investigate: (1) the transferability of our method under distribution shifts (i.e., cross-dataset evaluation);  and (2) the sensitivity of our approach to different risk definitions (i.e., varying $k$).


\subsection{Transferability Analysis} \label{appd:transferability}
To assess the generalization capability of \name to out-of-distribution tasks, we further examine whether the theoretical risk control holds under distribution shifts. Figure~\ref{fig:risknoniid} illustrates the distribution of admission risks on HumanEval using thresholds calibrated on MBPP.




Although the rigorous theoretical guarantee relies on the i.i.d. assumption, empirical results show that the admission risk remains largely controlled below the target tolerance $\alpha$. This indicates that \name is practically robust to moderate distribution shifts.




\input{tables/k5}

\subsection{Sensitivity to Risk Definition ($k=5$)} \label{appd:k5_analysis}
In the main experiments, we defined the risk based on the pass rate $H@k$ with $k=3$. To evaluate the sensitivity of \name to this hyperparameter, we conducted additional experiments setting $k=5$. This adjustment implies a slightly more relaxed criterion for success, effectively allowing the model more attempts to yield a correct solution.

We re-calibrated and evaluated \name under this new setting. As shown in Table~\ref{tab:abstention_k5}, the results  exhibit negligible variance compared to the $k=3$ case. \name continues to robustly identify and abstain from unsolvable tasks, maintaining its performance advantage over other methods. This consistency confirms that the effectiveness of \name is not dependent on a specific choice of $k$, demonstrating its robustness to different risk definitions.


%% file: tables/k5.tex
\begin{table*}[!t]
\centering
\small 
\caption{Task abstention results on HumanEval and MBPP with $k=5$. Our approaches (i.e., `SE+STDF' and `CR+STDF') generally outperform the competitors. }
\label{tab:abstention_k5}
\begin{tabularx}{\textwidth}{l | *{6}{>{\centering\arraybackslash}X}} 
\toprule
\multirow{2}{*}{\diagbox[width=6em, trim=l]{Method}{Model}} 
 & \multicolumn{2}{c}{DeepSeek-Coder} & \multicolumn{2}{c}{Qwen2.5-Coder}  & \multicolumn{2}{c}{WizardCoder} \\
 & \textbf{P} & \textbf{F1} & \textbf{P} & \textbf{F1}  & \textbf{P} & \textbf{F1} \\
\midrule

\multicolumn{7}{c}{\textbf{HumanEval}} \\
\midrule
PPL         & 36.41 & 53.39 & 17.79 & 30.20 & 39.42 & 52.23  \\
CLM         & 36.97 & 53.67 & 20.32 & 32.89 & 37.71 & 51.49  \\
CodeHalu    & 28.68 & 42.68 & 16.09 & 26.16 & 27.27 & 41.25  \\
SE          & 60.27 & 66.66 & 28.57 & 41.51 & 52.43 & \SOTA{63.70} \\
CR          & 59.49 & 68.11 & 41.02 & 47.06 & 51.19 & 62.77  \\
SE + STDF   & 56.71 & 64.95 & 46.66 & 50.91 & 53.33 & 62.50  \\
CR + STDF   & \SOTA{68.75} & \SOTA{71.54} & \SOTA{70.33} & \SOTA{53.90} & \SOTA{60.00} & 61.11 \\

\midrule

\multicolumn{7}{c}{\textbf{MBPP}} \\
\midrule
PPL         & 36.32 & 53.28 & 26.15 & 41.04 & 31.75 & 47.45  \\
CLM         & 36.32 & 53.28 & 31.60 & 44.25 & 35.10 & 50.00  \\
CodeHalu    & 33.43 & 48.63 & 28.12 & 41.02 & 32.74 & 47.32  \\
SE          & 51.12 & 64.60 & 34.80 & 45.65 & 50.00 & 61.26  \\
CR          & 59.18 & 66.28 & 31.95 & 45.57 & 60.99 & 63.70  \\
SE + STDF   & \SOTA{66.13} & \SOTA{72.88} & 46.76 & \SOTA{52.84} & 60.26 & \SOTA{64.99} \\
CR + STDF   & 59.39 & 66.47 & \SOTA{47.22} & 47.44 & \SOTA{62.59} & 63.08 \\

\bottomrule
\end{tabularx}
\end{table*}